%% file: sample-sigconf.tex
\documentclass[manuscript,screen]{acmart}
\AtBeginDocument{%
  }

\copyrightyear  { }
\acmYear{ }
\setcopyright{acmlicensed}\acmConference[ ]{ }{ }{ }
\acmBooktitle{ }
\acmPrice{15.00}
\acmDOI{10.1145/3543507.3583446}
\acmISBN{978-1-4503-9416-1/23/04}

\usepackage{multirow}
\usepackage{tikz}
\usepackage{enumitem}
\usepackage[utf8]{inputenc}
\usepackage{graphicx}
\usepackage{subfigure}
\usepackage{booktabs} 
\usepackage{pgfplots}
\usepackage{tikz}
\usepackage{amsmath}
\usepackage{amsfonts}

\usetikzlibrary{shapes.geometric, arrows, positioning}
\pgfplotsset{width=7cm,compat=1.17}

\tikzstyle{startstop} = [rectangle, rounded corners, minimum width=3cm, minimum height=1cm,text centered, draw=black, fill=red!30]
\tikzstyle{process} = [rectangle, minimum width=3cm, minimum height=1cm, text centered, draw=black, fill=orange!30]
\tikzstyle{arrow} = [thick,->,>=stealth]

\begin{document}
\settopmatter{printacmref=false}
\settopmatter{printccs=false}
\settopmatter{printfolios=false}
\settopmatter{authorsperrow=4}
\title{Towards Automated Model Design on Recommender Systems}

\author{Tunhou Zhang}
\authornote{A majority of this work was done when the first author was an intern at Meta Platforms, Inc.}
\affiliation{%
  \institution{Duke University}
  \city{Durham}
  \country{USA}
 } \email{tunhou.zhang@duke.edu}

\author{Dehua Cheng}
\affiliation{%
  \institution{Meta Platforms, Inc.}
  \city{Menlo Park}
  \country{USA}
}
\email{dehuacheng@meta.com}

\author{Yuchen He}
\affiliation{%
  \institution{Meta Platforms, Inc.}
  \city{Menlo Park}
  \country{USA}
}
\email{yuchenhe@meta.com}

\author{Zhengxing Chen}
\affiliation{%
  \institution{Meta Platforms, Inc.}
  \city{Menlo Park}
  \country{USA}
}
\email{czxttkl@meta.com}

\author{Xiaoliang Dai}
\affiliation{%
  \institution{Meta Platforms, Inc.}
  \city{Menlo Park}
  \country{USA}
}
\email{xiaoliangdai@meta.com}

\author{Liang Xiong}
\affiliation{%
  \institution{Meta Platforms, Inc.}
  \city{Menlo Park}
  \country{USA}
}
\email{lxiong@meta.com}

\author{Yudong Liu}
\affiliation{%
  \institution{Duke University}
  \city{Durham}
  \country{USA}
 }
 \email{yudong.liu@duke.edu}

\author{Feng Cheng}
\affiliation{%
  \institution{Duke University}
  \city{Durham}
  \country{USA}
 }
 \email{feng.cheng@duke.edu}

\author{Yufan Cao}
\affiliation{%
  \institution{Duke University}
  \city{Durham}
  \country{USA}
 }
 \email{yufan.cao@duke.edu}

\author{Feng Yan}
\affiliation{%
  \institution{University of Houston}
  \city{Houston}
  \country{USA}
}
\email{fyan5@central.uh.edu}

\author{Hai Li}
\affiliation{%
  \institution{Duke University}
  \city{Durham}
  \country{USA}
 }
 \email{hai.li@duke.edu}
 
\author{Yiran Chen}
\affiliation{%
  \institution{Duke University}
  \city{Durham}
  \country{USA}
 }
\email{yiran.chen@duke.edu}

\author{Wei Wen}
\authornote{Corresponding author. Intern Manager.}
\affiliation{
  \institution{Meta Platforms, Inc.}
  \city{Menlo Park}
  \country{USA}
}
\email{wewen@meta.com}

\begin{abstract}
The increasing popularity of deep learning models has created new opportunities for developing AI-based recommender systems. Designing recommender systems using deep neural networks requires careful architecture design, and further optimization demands extensive co-design efforts on jointly optimizing model architecture and hardware. Design automation, such as Automated Machine Learning (AutoML), is necessary to fully exploit the potential of recommender model design, including model choices and model-hardware co-design strategies.
We introduce a novel paradigm that utilizes weight sharing to explore abundant solution spaces. Our paradigm creates a large supernet to search for optimal architectures and co-design strategies to address the challenges of data multi-modality and heterogeneity in the recommendation domain. From a model perspective, the supernet includes a variety of operators, dense connectivity, and dimension search options. From a co-design perspective, it encompasses versatile Processing-In-Memory (PIM) configurations to produce hardware-efficient models.
Our solution space's scale, heterogeneity, and complexity pose several challenges, which we address by proposing various techniques for training and evaluating the supernet. Our crafted models show promising results on three Click-Through Rates (CTR) prediction benchmarks, outperforming both manually designed and AutoML-crafted models with state-of-the-art performance when focusing solely on architecture search. From a co-design perspective, we achieve $2\times$ FLOPs efficiency, 1.8$\times$ energy efficiency, and $1.5\times$ performance improvements in recommender models.

\end{abstract}




\maketitle

\noindent \textbf{Acknowledgement.}
Yiran Chen’s work is partially supported by the following grants:
NSF-2120333, NSF-2112562, NSF-1937435, NSF-2140247 and ARO W911NF-19-2-0107. Feng’s work is partially supported by the following grant: NSF CAREER-2048044. We also thank Maxim Naumov, Jeff Hwang, and Colin
Taylor in Meta Platforms, Inc., for their help with this project.

\input{_txt/1_Introduction}
\input{_txt/2_Related_Work}
\input{_txt/3_WS_NAS_Search_Space}
\input{_txt/4_WSNAS}

\input{_txt/5_Experiments}

\input{_txt/6_Codesign}
\input{_txt/7_Supplementary_Material}
\input{_txt/8_Conclusion}

\bibliographystyle{ACM-Reference-Format}
\bibliography{reference}

\end{document}


\settopmatter{printacmref=false}
\settopmatter{printccs=false}
\settopmatter{printfolios=false}
\settopmatter{authorsperrow=5}
\title{NASRec: Weight Sharing Neural Architecture Search for Recommender Systems}

\maketitle

\setcounter{section}{6}
\setcounter{table}{6}
\setcounter{figure}{4}

\input{_txt/Supplementary_Material.tex}

%% file: _txt/1_Introduction.tex
\section{Introduction}

\begin{table*}[b]
\vspace{-1em}
\caption{Comparison of our approach vs. existing AutoML methods for recommender systems, from a \textbf{model} perspective.}
\begin{center}
\scalebox{1.0}{
    \begin{tabular}{|c|c|c|c|c|c|}
        \hline
        \multirow{2}{*}{\textbf{Method}} & \textbf{Building} & \textbf{Dense} & \textbf{Full arch} & \textbf{Co-design} & \textbf{Criteo} \\
         & \textbf{Operators?} & \textbf{Connectivity?} & \textbf{Search?} & \textbf{Support?} & \textbf{Log Loss} \\
    \hline
    \textbf{DNAS}~\cite{krishna2021differentiable} & FC, Dot-Product & \checkmark & & & 0.4442 \\
    \textbf{PROFIT}~\cite{gao2021progressive} & FC, FM & &  & & 0.4427 \\
    \textbf{AutoCTR}~\cite{song2020towards} & FC, Dot-Product, FM, EFC & \checkmark & \checkmark & & 0.4413 \\
    \multirow{2}{*}{\textbf{Ours}} & FC, Gating, Sum, Attention, & \multirow{2}{*}{\checkmark} & \multirow{2}{*}{\checkmark} & \multirow{2}{*}{\checkmark} & \multirow{2}{*}{\textbf{0.4399}} \\ 
     & Dot-Product, FM, EFC & & & & \\
    \hline
    \end{tabular}    
}
\end{center}
\label{tab:intro}
\end{table*}

Recommender systems, which are widely used in search engines and social media platforms~\cite{kowald2017temporal,carterette2007evaluating} to optimize Click-Through Rates (CTR), rely on deep learning-based models~\cite{covington2016deep, naumov2019deep, guo2015trustsvd} that incorporate multi-modality features. However, these models present challenges in feature interaction modeling and neural network optimization due to the heterogeneity of the features. Finding a good backbone model with appropriate priors on multi-modality features is standard practice, but it requires significant manual efforts~\cite{rendle2011fast,richardson2007predicting,he2014practical,cheng2016wide, shan2016deep,guo2017deepfm,lian2018xdeepfm,naumov2019deep} and is limited by available resources.

Automated Machine Learning (AutoML) techniques, such as Weight-Sharing Neural Architecture Search (WS-NAS)~\cite{liu2018darts,yu2020bignas,cai2019once}, have shown promise in optimizing the design of efficient models for recommender systems without human intervention. 
NAS seeks the best architecture choice within an abundant solution space, with versatile search strategies~\cite{krishna2021differentiable,zoph2018learning,mellor2021neural} and evaluation strategies of models.
However, these techniques face unique challenges due to the multi-modality and heterogeneity of data and architecture in recommendation systems compared to vision models.
One challenge is that the inputs to building blocks in recommendation systems are multi-modal and generate 2D and 3D tensors, whereas vision models have homogeneous 3D tensors~\cite{cai2019once,wang2020hat} as inputs. Additionally, while state-of-the-art NAS in vision converges to searching size configurations, recommendation models are heterogeneous, with each stage using a different building block. Furthermore, recommendation models use a variety of heterogenous operators, such as Fully-Connected layer~\cite{cheng2016wide}, Dot-Product~\cite{naumov2019deep}, Multi-Head Attention~\cite{song2019autoint}, and Factorization Machine~\cite{guo2017deepfm,lian2018xdeepfm}. In contrast, vision models mainly use homogenous convolutional operators and backbone search on design motifs~\cite{cai2019once,yu2020bignas}, such as layer width, layer depth, and kernel size.
These challenges are further complicated by co-design tasks such as mixed-precision quantization search, which can worsen the quality of the product driven by AutoML. Overall, there is a need for more effective AutoML techniques that can handle the unique challenges of data multi-modality and architecture heterogeneity in recommendation systems.

Due to the challenges above, the study of AutoML in recommender systems is limited.
For example, search spaces in AutoCTR~\cite{song2020towards} and DNAS~\cite{krishna2021differentiable} follow the design principle of human-crafted DLRM~\cite{naumov2019deep}, and they only include Fully-Connected layer and Dot-Product as searchable operators. They also heavily rely on manually crafted operators, such as Factorization Machine~\cite{song2020towards} or feature interaction module~\cite{gao2021progressive} in the search space to increase architecture heterogeneity.
Moreover, existing works suffer from huge computation costs~\cite {song2020towards} or challenging bi-level optimization~\cite{krishna2021differentiable}. Thus, they only employ narrow design spaces (sometimes with strong human priors~\cite{gao2021progressive}) to craft architectures, discouraging diversified feature interactions and harming the quality of crafted models.

We propose a full-stack solution paradigm to fully enable Neural Architecture and Parameter Search via WS-NAS under data modality and architecture heterogeneity. We offer end-to-end solutions on model design and hardware co-design strategies.
From a model perspective, we summarize the advancement of our proposed paradigm over other NAS approaches in Table~\ref{tab:intro}. 
We achieve this by building a supernet incorporating much more heterogeneous operators than previous works, including a Fully Connected (FC) layer, Gate, Sum, Dot-Product, Self-Attention, and Embedded Fully Connected (EFC) layer.
In the supernet, we densely connect a cascade of blocks, each including all operators as options. 
As dense connectivity allows any block to take in any raw feature embeddings and intermediate tensors, the supernet is not limited by any particular data modality.
Such supernet design minimizes the encoding of human priors~\cite{zhang2023nasrec}, supporting the nature of data modality and architecture heterogeneity in recommenders and covering models beyond popular recommendation models
such as Wide \& Deep~\cite{cheng2016wide}, DeepFM~\cite{guo2017deepfm}, DLRM~\cite{naumov2019deep}, AutoCTR~\cite{song2020towards}, DNAS~\cite{krishna2021differentiable}, and PROFIT~\cite{gao2021progressive}.
Our approach also supports the co-design of model and hardware (e.g., Processing-In-Memory architecture~\cite{yang2022research}), providing headroom for improvement when deploying recommender models in reality. The ad-hoc analysis of structured pruning on our crafted models further expands the opportunity to improve the efficiency of recommender models.

The supernet essentially forms a search space. We obtain a model by zeroing out some operators and connections in the supernet; that is, a subnet of the supernet is equivalent to a model.
All subnets share weights from the same supernet called Weight Sharing NAS (WS-NAS).
To efficiently search models/subnets in the search space, we advance one-shot approaches~\cite{cai2019once,yu2020bignas} to the recommendation domain.
We propose \textit{Single-operator Any-connection sampling} to decouple operator selections and increase connection coverage,
\textit{operator-balancing interaction} blocks to train subnets in the supernet fairly, and \textit{post-training fine-tuning} to reduce weight co-adaptation.
These approaches enable a better training efficiency and ranking of subnet models in the supernet, resulting in $\sim$0.001 log loss reduction of searched models on full NASRec search space.
We further apply the search on model-hardware co-design and study ad hoc structured pruning, unlocking extra benefits in the efficiency of recommender models.

From a model perspective, we evaluate our AutoML-crafted models on three popular CTR benchmarks and demonstrate significant improvements compared to hand-crafted and NAS-crafted models.
Remarkably, our approach advances the state-of-the-art with log loss reduction of $\sim 0.001$ and $\sim 0.003$ on Criteo and KDD Cup 2012, respectively.
On Avazu, our approach advances the state-of-the-art PROFIT~\cite{gao2021progressive} with AUC improvement of $\sim 0.002$ and on-par log loss while outperforming PROFIT~\cite{gao2021progressive} on Criteo by $\sim$0.003 log loss reduction.
Thanks to the efficient weight-sharing mechanism, our approach only needs to train a single supernet, greatly reducing the search cost. 
From a co-design perspective, we offer a detailed analysis to exploit the potential of our crafted models and uncover 1.5$
\times$ theoretical speedup on discovered models.
The ad-hoc structured pruning achieves $\sim$2$\times$ FLOPs saving without harming log loss and AUC on recommender benchmarks.

We demonstrate the outline of our manuscript as follows.
Section \ref{sec:related_work} introduces the related work in recommender systems.
Section \ref{sec:nasrec_space} elaborates on the search space from both the model and co-design perspectives. 
In Section \ref{sec:nasrec_method}, we propose the search methodology and demonstrate the main technologies.
In Section \ref{sec:nasrec_exps}, we evaluate our crafted models on 3 CTR benchmarks, demonstrating state-of-the-art performance and uncovering theoretical efficiency gain from model-hardware co-design.
In Section \ref{sec:nasrec_ablation}, we provide ablation studies and discussions to better understand our system and methodologies, including structured pruning to advance the model efficiency.
In Section \ref{sec:conclusion}, we present our conclusion.
We summarize our major contributions below.
\begin{itemize}[noitemsep,leftmargin=*]
    \item We propose a new paradigm to scale up the automated design of recommender systems from both the model and co-design perspectives.
    NASRec establishes a flexible supernet (search space) with minimal human priors, overcoming data modality and architecture heterogeneity challenges in the recommendation domain.
    \item We advance weight-sharing NAS to the recommendation domain by introducing single-operator any-connection sampling, operator-balancing interaction modules, and post-training fine-tuning.
    \item From a model perspective, our crafted models outperform hand-crafted and AutoML-crafted models with a smaller search cost.
    \item From a co-design perspective, we explore various choices to co-design model architecture with Processing-In-Memory hardware, demonstrating significant speed-up headroom.
\end{itemize}

%% file: _txt/2_Related_Work.tex
\section{Related Work}
\label{sec:related_work}

\noindent \textbf{Deep learning based recommender systems.}
Machine-based recommender systems, such as those predicting Click-Through Rates (CTR), have been extensively studied using various approaches like Logistic Regression~\cite{richardson2007predicting}, Gradient-Boosting Decision Trees~\cite{he2014practical}, Wide \& Deep Neural Networks~\cite{cheng2016wide}, crossing networks~\cite{shan2016deep}, Factorization Machines~\cite{guo2017deepfm,lian2018xdeepfm}, Dot-Product~\cite{naumov2019deep}, and gating mechanisms~\cite{wang2017deep, wang2021dcn,yan2023xdeepint}. 
Additionally, researchers have explored efficient feature interactions via feature-wise multiplications~\cite{wang2021masknet} and sparsifications~\cite{deng2021deeplight} to develop lightweight recommender systems. However, these methods require significant manual efforts and may result in suboptimal performance due to limited resource availability and constrained design choices.
Our work introduces a novel paradigm for learning effective recommender models, including novel model architecture search space and effective model hardware co-design via Processing-In-Memory hardware and mixed-precision quantization.

\noindent \textbf{AutoML and NAS.}     
Automated Machine Learning (AutoML) has gained significant popularity in automating the design of Deep Neural Networks across various applications such as Computer Vision~\cite{zoph2018learning,liu2018darts,wen2020neural,cai2019once,Wang2020APQ}, Natural Language Processing~\cite{so2019evolved,wang2020hat}, and Recommendation Systems~\cite{song2020towards,gao2021progressive,krishna2021differentiable}. 
Neural Architecture Search (NAS), especially Weight-Sharing Neural Architecture Search (WS-NAS)~\cite{cai2019once,wang2020hat}, has recently garnered attention due to its ability to train a supernet representing the entire search space directly on target tasks and efficiently evaluate subnets with shared supernet weights. However, applying WS-NAS to recommender systems is challenging because these systems involve heterogeneous architectures dedicated to interacting with multi-modality data, requiring more flexible search spaces and effective supernet training algorithms.
These challenges lead to co-adaptation~\cite{bender2018understanding} and operator-imbalance problems~\cite{liang2019darts+} in WS-NAS, resulting in lower rank correlation for distinguishing models. To address these issues, our work introduces a series of technical solutions: single-operator any-connection sampling, operator-balancing interaction modules, and post-training fine-tuning to address these challenges.
In addition, our work considers both joint architecture-hardware search and ad-hoc mixed-precision exploration to enhance discovered models, providing novel perspectives and insights on recommender system model designs.

\noindent \textbf{Software-hardware Co-design.}
Classic software-hardware co-design works mainly focus on the joint optimization of the model architecture and hardware execution throughput~\cite{cai2019once,wang2020hat}.
In this work, we explore Processing-in-memory (PIM) architectures and discuss the co-design of recommender models with PIM architectures for real-world applications, and connect PIM optimization with model compression techniques such as pruning and quantization.
PIM uses crossbar-based structures in advanced memory technologies such as Resistive Random-Access Memory (ReRAM)~\cite{yang2022research}.
Prior research in this field~\cite{wang2021rerec} has explored using PIM's inherent parallel processing capabilities to enhance the performance of recommender systems. 
However, many of these studies have not fully addressed the unique challenges posed by PIM-based recommender systems, such as uneven cache access patterns, inefficient mapping strategies, and sub-optimal heuristic-based design methodologies.
Further exploration in this field~\cite{gibbon} reveals that varying configurations of PIM—such as the size of crossbars, the precision of Analog-to-Digital Converters (ADCs) and Digital-to-Analog Converters (DACs), and the resolution of crossbars—can significantly influence key performance metrics like accuracy and energy consumption.
Our work incorporates the optimization of PIM design into model architecture search and demonstrates the initiative to craft hardware-friendly models for recommendation and user personalization.

%% file: _txt/3_WS_NAS_Search_Space.tex
\section{Hierarchical Search Space for Model and Co-design}
\label{sec:nasrec_space}
\begin{figure*}[t]
    \begin{center}
    \includegraphics[width=0.75\linewidth]{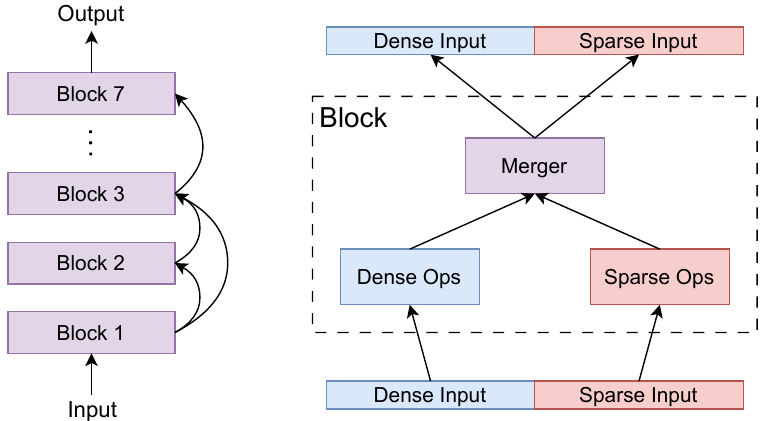}
    \label{fig:wsnas_search_space}
    \caption{NASRec search space enables a full architecture search on dense connectivity of blocks, dense/sparse operators, and mergers that fuse dense/sparse representations.}
    \end{center}
\end{figure*}

We first introduce the hierarchical search space design dedicated to architecture search from a model perspective.
We propose our resolutions to craft more hardware-friendly models from a co-design perspective.
This complements our vision to start AutoML research and realize an end-to-end recommender model design paradigm for social good.

\subsection{Model Architecture Search Space}
\label{sec:search_space}
Thus, we first demonstrate the model architecture space design by revisiting NASRec~\cite{zhang2023nasrec}.
The flexibility of search space is the key to supporting data modality and architecture heterogeneity in recommender systems.
The major manual process in designing the search space is simply collecting common operators used in existing approaches~\cite{guo2017deepfm,lian2018xdeepfm, naumov2019deep, song2020towards,wang2017deep, wang2021dcn}.
Beyond that, we further incorporate the prevailing Transformer Encoder~\cite{vaswani2017attention} into the search space for better flexibility and higher potential in searched architectures, thanks to its dominance in applications such as ViT~\cite{dosovitskiy2020image} for image recognition, Transformer~\cite{vaswani2017attention} for natural language processing, and its emerging exploration in recommender systems~\cite{chen2019behavior, gao2020modularized}.

In recommender systems, we define a dense input as $X_{d} \in \mathbb{R}^{B \times dim_d}$ which is a 2D tensor from either raw dense features or generated by operators, such as FC, Gating, Sum, and Dot-Product.
A sparse input $X_{s} \in \mathbb{R}^{B \times N_s \times dim_s}$ is a 3D tensor of sparse embeddings either generated by raw sparse/categorical features or by operators such as EFC and self-attention.
Similarly, a dense or sparse output (i.e., $Y_{d}$ or $Y_{s}$) is respectively defined as a 2D or 3D tensor produced via corresponding building blocks/operators.
In NASRec, all sparse inputs and outputs share the same $dim_s$, which equals to the dimension of raw sparse embeddings.
Accordingly, we define a dense (sparse) operator as one that produces a dense (sparse) output. 
In NASRec, dense operators include FC, Gating, Sum, and Dot-Product, which form the ``dense branch''; sparse operators include EFC and self-attention, which form the ``sparse branch''. 

A candidate architecture in NASRec search space is a stack of $N$ choice blocks, followed by a final FC layer to compute the final logit.
Each choice block admits an arbitrary number of multi-modality inputs, each of which is $X=(X_{d}, X_{s})$ from a previous block or raw inputs, and produces a multi-modality output $Y=(Y_{d}, Y_{s})$ of both a dense tensor $Y_{d}$ and a sparse tensor $Y_{s}$ via internal building operators.
Within each choice block, we can sample operators for search.

We construct a supernet representing the NASRec search space; see Figure \ref{fig:wsnas_search_space}.
The supernet subsumes all possible candidate models/subnets and performs weight sharing among subnets to train them simultaneously.
We formally define the NASRec supernet $\mathcal{S}$ as a tuple of connections $\mathcal{C}$, operators $\mathcal{O}$, and dimensions $\mathcal{D}$ as follows: $\mathcal{S} = (\mathcal{C}, \mathcal{D}, \mathcal{O})$ over all $N$ choice blocks. Specifically, the operators: $\mathcal{O}=[O^{(1)}, ..., O^{(N)}]$ enumerates the set of building operators from choice block $1$ to $N$.
The connections: $\mathcal{C}=[C^{(1)}, ..., C^{(N)}]$ contains the connectivity $<i, j>$ between choice block $i$ and choice block $j$.
The dimension: $\mathcal{D}=[D^{(1)}, ..., D^{(N)}]$ contains the dimension settings from choice block $1$ to $N$.

A subnet $S_{sample}=(\mathcal{O}_{sample}, \mathcal{C}_{sample}, \mathcal{D}_{sample})$ in the supernet $\mathcal{S}$ represents a model in NASRec search space. 
A block uses addition to aggregate the outputs of sampled operators in each branch (i.e., ``dense branch'' or ``sparse branch''). When the operator output dimensions do not match, we apply zero masking to mask the extra dimension.
A block uses concatenation $Concat$
to aggregate the outputs from sampled connections.
Given a sampled subnet $S_{sample}$, the input $X^{(N)}$ to choice block $N$ is computed as follows given a list of previous block outputs $\{Y^{(1)}, ..., Y^{(N-1)}\}$ and the sampled connections $C_{sample}^{(N)}$:

\begin{equation}
    X_d^{(N)} = Concat_{i=1}^{N-1}[Y_d^{(i)} \cdot \mathbf{1}_{<i, N> \in C^{(N)}_{sample}}],    
\end{equation}
\begin{equation}
    X_s^{(N)} = Concat_{i=1}^{N-1}[Y_s^{(i)} \cdot \mathbf{1}_{<i, N> \in C^{(N)}_{sample}}].
\end{equation}

Here, $\mathbf{1}_b$ is 1 when $b$ is true otherwise 0.

A building operator $o \in O^{(N)}_{sample}$ transforms the concatenated input $X^{(N)}$ into an intermediate output with a sampled dimension $D^{(N)}_{sample}$. This is achieved by applying a mask function on the last dimension for dense output and the middle dimension for sparse output.
For example, a dense output $Y^{(N)}_{d}$ is obtained as follows:

\begin{equation}
    Y_d^{(N)} = \sum_{o \in \mathcal{O}} {\mathbf{1}_{o \in \mathcal{O}_{sample}^{(N)}} \cdot Mask(o(X_d^{(N)}), D^{(N)}_{sample, o})}.
\end{equation}
where
\begin{equation}
\label{eq:mask_dim}
    Mask(V, d) = \begin{cases}
V_{:, i}, \text{ if } i<d \\
0, \text{ Otherwise.}\\
\end{cases}.
\end{equation}

Next, we clarify the set of dense/sparse building operators as follows:
\begin{itemize}[noitemsep,leftmargin=*]
    \item \textbf{Fully-Connected (FC) layer.} The connected layer is the backbone of DNN models for recommender systems~\cite{cheng2016wide} that extracts dense representations. FC is applied on 2D dense inputs, and followed by a ReLU activation. 
    
    \item \textbf{Sigmoid Gating (SG) layer.} 
    We follow the intuition in ~\cite{wang2021dcn,chen2019behavior} and employ a dense building operator, Sigmoid Gating, to enhance the potential of the search space. 
    Given two dense inputs $X_{d1}\in \mathbb{R}^{B \times dim_{d1}}$ and $X_{d2}\in \mathbb{R}^{B \times dim_{d2}}$, Sigmoid Gating interacts these two inputs as follows: $SG(X_{d1}, X_{d2}) = sigmoid(FC(X_{d1})) * X_{d2}$.
    If the dimensions of two dense inputs do not match, zero padding is applied to the input with the lower dimension.
        
    \item \textbf{Sum layer.} This dense building operator adds two dense inputs: $X_{d1} \in \mathbb{R}^{B \times dim_{d1}}$, $X_{d2} \in \mathbb{R}^{B \times dim_{d2}}$ and merges two features from different levels of the recommender system models by simply performing $Sum(X_{d1}, X_{d2})=X_{d1} + X_{d2}$.
    Like Sigmoid Gating, zero padding is applied on the input with a lower dimension.

    \item \textbf{Dot-Product (DP) layer.} We leverage Dot-Product to grasp the interactions among multi-modality inputs via a pairwise inner product.
    Dot-Product can take dense and/or sparse inputs and produce a dense output.
    After being sent to the ``dense branch,'' these sparse inputs can later use the dense operators to learn better representations and interactions.
    Given a dense input $X_{d} \in \mathbb{R}^{B \times dim_{d}}$ and a sparse input $X_{s} \in \mathbb{R}^{B \times N_c \times dim_{s}}$, a Dot-Product first concatenate them as $X = Concat[X_d, X_s]$, and then performs pair-wise inner products: $DP(X_d, X_s)=Triu(XX^{T})$. $dim_{d}$ is first projected to $dim_{s}$ if they do not match.

    \item \textbf{Embedded Fully-Connected (EFC) layer}. An EFC layer is a sparse building operator that applies FC along the middle dimension. Specifically, an EFC with weights $W \in \mathbb{R}^{N_{in} \times N_{out}}$ transforms an input $X_s \in \mathbb{R}^{B \times N_{in} \times dim_{s}}$ to $Y_s \in \mathbb{R}^{B \times N_{out} \times dim_{s}}$
    
    \item \textbf{Attention (Attn) layer.} Attention layer is a sparse building operator that utilizes the Multi-Head Attention (MHA) mechanism to learn the weighting of sparse inputs and better exploit their interaction in recommendation systems. Here, We apply Transformer Encoder on a given sparse input $X_{s} \in \mathbb{R}^{B\times N_s \times dim_{s}}$, with identical queries, keys, and values.
\end{itemize}

We observe that the aforementioned set of building operators provides opportunities for the sparse inputs to transform into the ``dense branch''.
Yet, these operators do not permit a transformation of dense inputs towards the ``sparse branch''.
To address this limitation,
we introduce "\textbf{dense-sparse merger}"
allow dense/sparse outputs to merge into the ``sparse/dense branch optionally''. Dense-sparse merger contains two major components.
\begin{itemize}[noitemsep,leftmargin=*]
    \item "Dense-to-sparse" merger. This merger first projects the dense outputs $X_{d}$ using an FC layer, then uses a reshape layer to reshape the projection into a 3D sparse tensor. The reshaped 3D tensor is merged into the sparse output via concatenation. 
    \item "Sparse-to-dense" merger. This merger employs a Factorization Machine (FM)~\cite{guo2017deepfm} to convert the sparse output into a dense representation and then add the dense representation to the dense output. 
\end{itemize}

Beyond the rich choices of building operators and mergers, each choice block can receive inputs from any preceding choice blocks, and raw input features. 
This involves exploring any connectivity among choice blocks and raw inputs, extending the wiring heterogeneity for search.

\subsection{Co-design Model Architecture and Hardware}
\begin{figure}[!h]
    \begin{center}
    \includegraphics[width=0.75\linewidth]{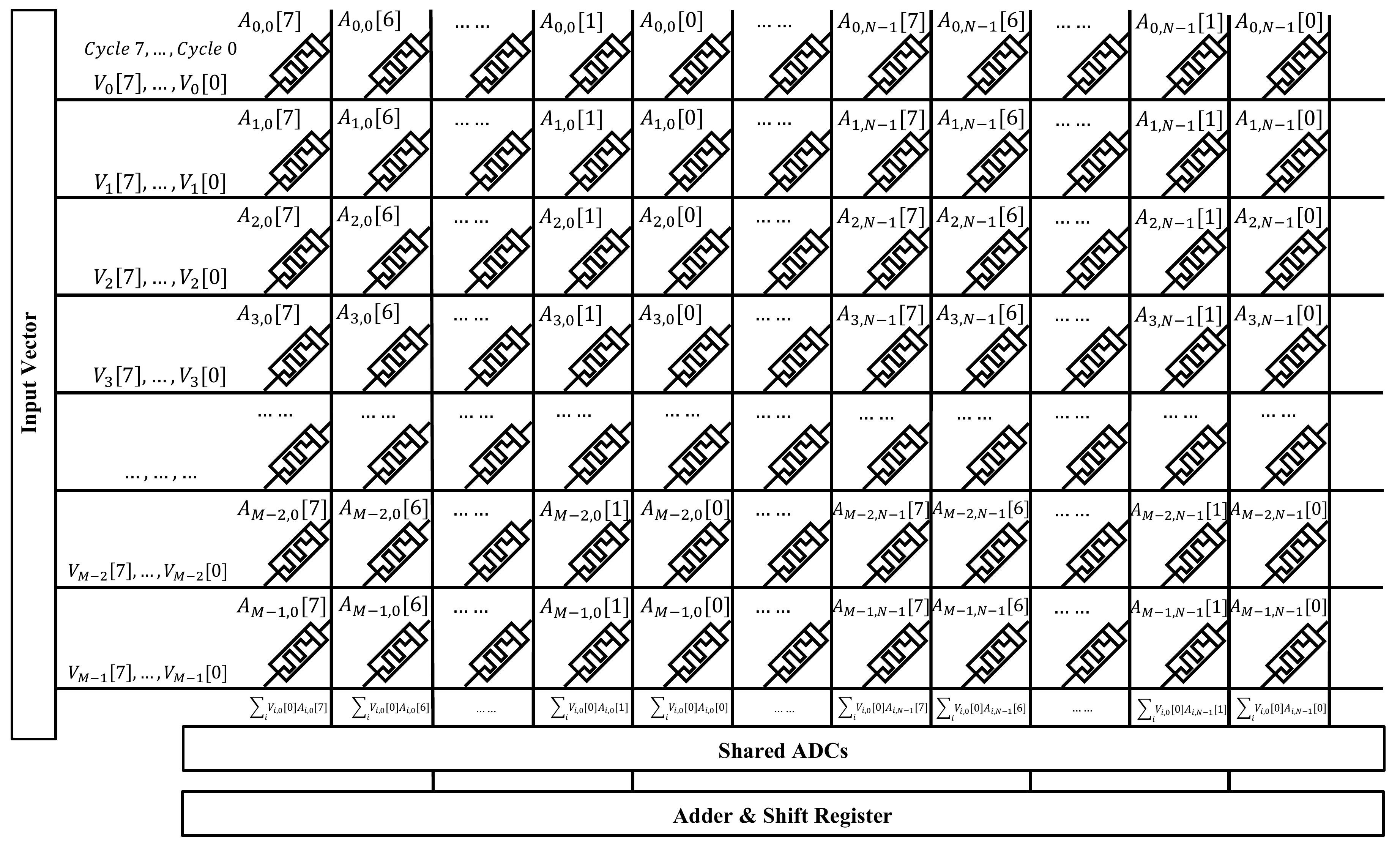}
    \caption{Operation principle of ReRAM-based in-memory computing.}
    \label{fig:xb}
    \end{center}
\end{figure}

Next, we consider further enhancing the co-design of model architecture and hardware, specifically, 
Figure \ref{fig:xb} illustrates an overview of PIM hardware design, representing a significant innovation in computing. These designs apply analog voltages to each Word Line (WL), initiating a process where these voltages are multiplied by the conductance present in each row. This multiplication adheres to the principles of Ohm’s Law. Following this, the currents produced from this multiplication are combined along each column according to Kirchhoff’s Current Law. At the end of each Bit Line (BL), specialized circuitry interprets these aggregated currents to facilitate complex Matrix-Vector Multiplication (MVM) functions within the memory array.

As such, we propose integrating ReRAM-related parameters into the search space from a co-design perspective. This enables simultaneous and efficient co-exploration of the recommender system model architecture and the Processing-In-Memory (PIM) architecture. 
We map the building above operators onto the crossbars with minimal effort for straightforward evaluation.
Operators like EmbedFC, FC, and the dense-to-sparse merger are intrinsically MVM and follow the mapping protocol outlined in the background section. DP and sparse-to-dense-merger (i.e., Factorization Machine~\cite{guo2017deepfm}), not ideal for PIM, are assigned to the digital functional unit instead.

Our vision adopts mixed-precision search as a preliminary study for PIM hardware co-design, such as crossbar optimization and Resistive Random-Access Memory (ReRAM) optimization.
This is because quantization provides headroom analysis in ReRAM design towards hardware metrics such as digital-to-analog converter (DAC) resolution, Memristor precision, analog-to-digital converter (ADC) resolution, etc.
We provide theoretical analysis and demonstrate the simulation results on hardware, providing concrete guidance on the theoretical headroom of co-designing software and hardware for recommender models.

\subsection{Search Components} 
In the NASRec search space dedicated to model architectures, we search for each choice block's connectivity, operator dimensions, and building operators.
We illustrate the three key search components as follows:

\begin{itemize}[noitemsep,leftmargin=*]
\item \textbf{Connection.}
We place no restrictions on the number of connections a choice block can receive: each block can choose inputs from an arbitrary number of preceding blocks and raw inputs. 
Specifically, the n-th choice block can connect to any previous $n-1$ choice blocks and the raw dense (sparse) features.
The outputs from all preceding blocks are concatenated as inputs for dense (sparse) building blocks.
We separately concatenate the dense (sparse) outputs from preceding blocks. 

\item \noindent \textbf{Dimension.}
In a choice block, different operators may produce different tensor dimensions.
In NASRec, we set the output sizes of FC and EFC to $dim_{d}$ and $N_{s}$, respectively, and other operator outputs in the dense (sparse) branch are linearly projected to $dim_{d}$ ($N_{s}$).
This ensures operator outputs in each branch have the same dimension and can be added together. This also give the maximum dimensions $dim_{d}$ and $N_{s}$ for the dense output $Y_{d} \in \mathbb{R}^{B \times dim_d}$ and the sparse output $Y_{s} \in \mathbb{R}^{B \times N_s \times dim_s}$.
Given a dense or sparse output, a mask in Eq.~\ref{eq:mask_dim} zeros out the extra dimensions, allowing a flexible selection of building operators' dimensions.

\item \textbf{Operator.} Each block can choose at least one dense (sparse) building operator to transform inputs to a dense (sparse) output. Each block should maintain at least one operator in the dense (sparse) branch to ensure the flow of information from inputs to logit.
We independently sample building operators in the dense (sparse) branch to form a validated candidate architecture.
In addition, we independently sample dense-sparse mergers to allow optional dense-to-sparse interaction.

\end{itemize}

We showcase two model architecture search spaces as examples.

\begin{itemize}[noitemsep,leftmargin=*]
    \item \textit{NASRec-Small.} We limit the choice of operators within each block to FC, EFC, and Dot-Product and allow any connectivity between blocks. This provides a similar scale of search space as AutoCTR~\cite{song2020towards}.   

    \item \textit{NASRec-Full.} We enable all building operators, mergers, and connections to construct an aggressive search space for exploration with minimal human priors. Under the constraint that at least one operator must be sampled in both dense and sparse branches, the \textit{NASRec-Full} search space size is $15^{N}\times$ of \textit{NASRec-Small}, where $N$ is the number of choice blocks. This full search space extremely tests the capability of NASRec.
    \end{itemize}

The combination of full dense connectivity search and independent dense/sparse dimension configuration gives the model architecture search space a large cardinality.
\textit{NASRec-Full} has $N=7$ blocks, containing up to $5\times 10^{33}$ architectures with strong heterogeneity.
With minimal human priors and such unconstrained search space, brutal-force sample-based methods may take enormous time to find a state-of-the-art model.

In addition, we construct the co-design search space as follows:
\begin{itemize}[noitemsep,leftmargin=*]
    \item \textbf{DNN Design Space.} The DNN design space follows \textit{NASRec-Small} search space dependent on the compatibility of building operators on PIM hardware. This includes dense operators like FC and DP, with feature dimensions ranging from 64 to 1024. We also incorporate sparse operators with dimensions from 16 to 64 and dense-sparse interaction operators, including FC and FM.
    \item \textbf{Quantization Design Space.} We allow mapping onto all previously mentioned operators, including FC and EFC layers and FC and EFC projections inside DP and FM, but excluding DP and FM, as they are not a natural fit for ReRAM. The quantization of weights ranges from 4 to 8 bits.
\end{itemize}

%% file: _txt/4_WSNAS.tex
\section{Weight sharing Neural Architecture Search for Recommender Systems}
\label{sec:nasrec_method}
A NASRec supernet simultaneously breeds different subnet models in the aforementioned model and co-design search space, yet its large cardinality challenges training efficiency and ranking quality.
This section proposes a novel path sampling strategy, \textit{Single-operator Any-connection} sampling, that combines operator sampling with a good connection sampling coverage.
We further observe the operator imbalance phenomenon induced by some over-parameterized operators and tackle this issue by \textit{operator-balancing interaction} to improve supernet ranking.
Finally, we employ \textit{post-training fine-tuning} to alleviate weight co-adaptation and utilize regularized evolution to obtain the best subnet.
We also provide insights that effectively explore the best recommender models.

\begin{figure}[t]
    \begin{center}
    \includegraphics[width=0.75\linewidth]{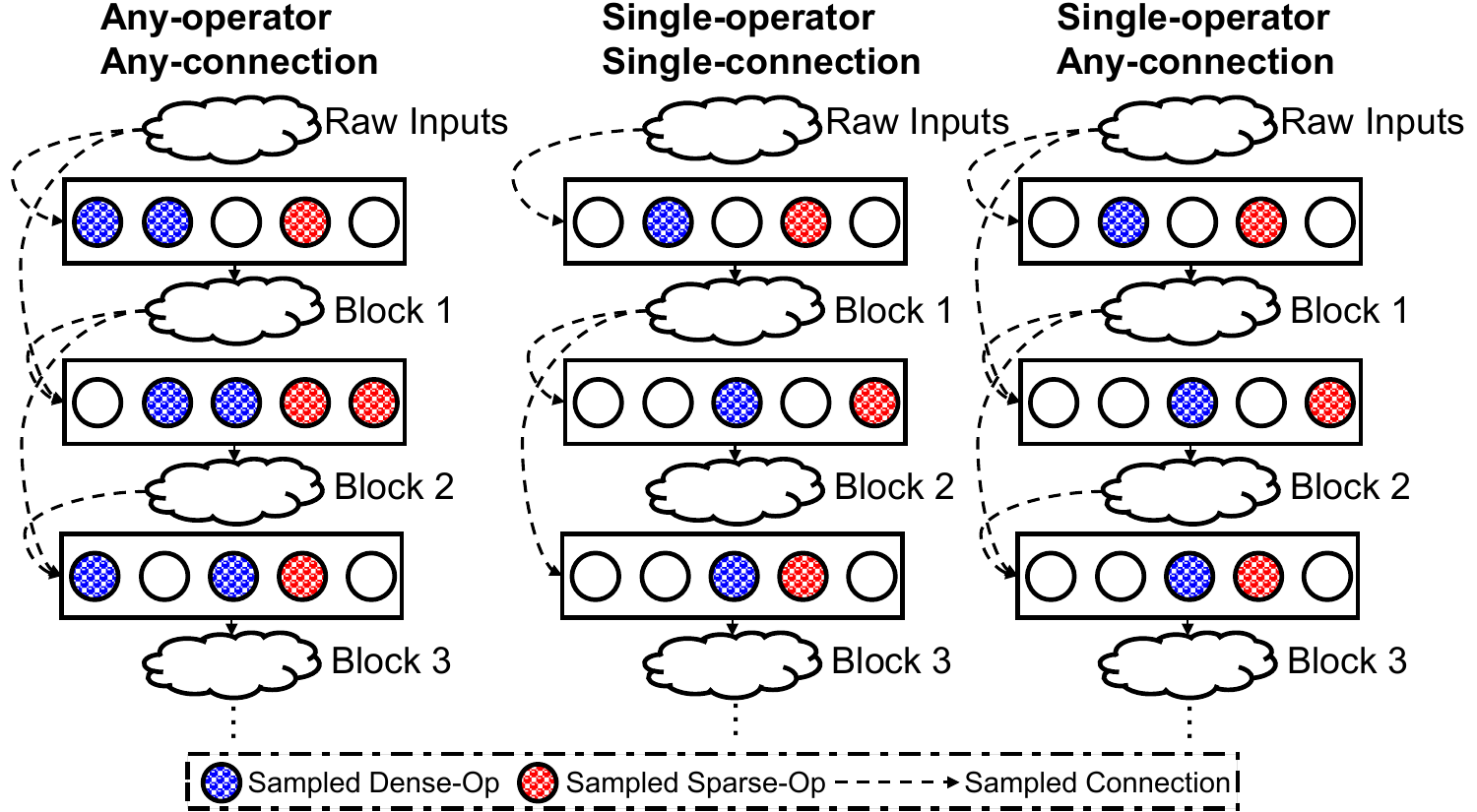}
    \caption{Single-operator Any-connection path sampling combines the advantages of the first two sampling strategies.}
    \label{fig:single_path}    
    \end{center}
\end{figure}

\subsection{Single-operator Any-Connection Sampling}
\label{sec:single_path}
During supernet training, a drop-out-like approach is adopted where, at each mini-batch, a subnet is sampled and trained. The goal is to train subnets that can well predict the performance of models under weight sharing. The sampling strategy used is critical to achieve this goal. Three path sampling strategies have been explored, and Single-operator Any-Connection sampling is the most effective among them:

\noindent \textbf{Single-operator Single-connection strategy}: This path sampling strategy, which has its roots in Computer Vision, uniformly samples a single dense and sparse operator in each choice block and a single connection as input to a block. While this strategy is efficient because it trains only a small subnet at a time, it encourages only chain-like formulations of models without extra connectivity patterns, leading to slower convergence, poor performance, and inaccurate ranking of models.

\noindent \textbf{Any-operator Any-connection Strategy}: This sampling strategy increases the coverage of sub-architectures of the supernet during subnet training by uniformly sampling an arbitrary number of dense and sparse operators in each choice block and an arbitrary number of connections to aggregate different block outputs. However, training efficiency is poor when training large subnets sampled in this way. Moreover, the co-adaptation of multiple operators within a choice block may affect the independent evaluation of subnets and lead to poor ranking quality.

\noindent \textbf{Single-operator Any-connection}: This path sampling strategy combines the strengths of the first two strategies. It samples a single dense and a single sparse operator in each choice block while allowing the sampling of an arbitrary number of connections to aggregate outputs from different choice blocks. The key insight behind this strategy is to separate the sampling of parametric operators to avoid weight co-adaptation while allowing the sampling of non-parametric connections to gain good coverage of the search space.

Here, dashed connections and operators denote a sampled path in the supernet.
Compared to Any-operator Any-connection sampling, 
single-operator Any-connection sampling achieves higher training efficiency: the reduced number of sampled operators reduces the training cost by up to 1.5$\times$.
In addition, Single-operator Any-connection samples medium-sized networks more frequently.
These medium-sized networks achieve the best trade-off between model size and performance, as shown in Table ~\ref{tab:model_complexity}.

\subsection{Operator-Balancing Interaction Modules}
Recommender systems involve multi-modality data with an indefinite number of inputs, such as many sparse inputs.
We define operator imbalance as the imbalance of the number of weights between operators within a block.
In weight-sharing NAS, operator imbalance may cause supernet training to favor operators with more weights. This will offset the gains due to poor ranking correlations of subnets: the subnet performance in the supernet may deviate from its ground-truth performance when trained from scratch.
Within the NASRec search space, we identify that such an issue is strongly related to the Dot-Product operator and provide mitigation to address such operator imbalance.

\begin{figure}[t]
    \begin{center}
    \includegraphics[width=0.7\linewidth]{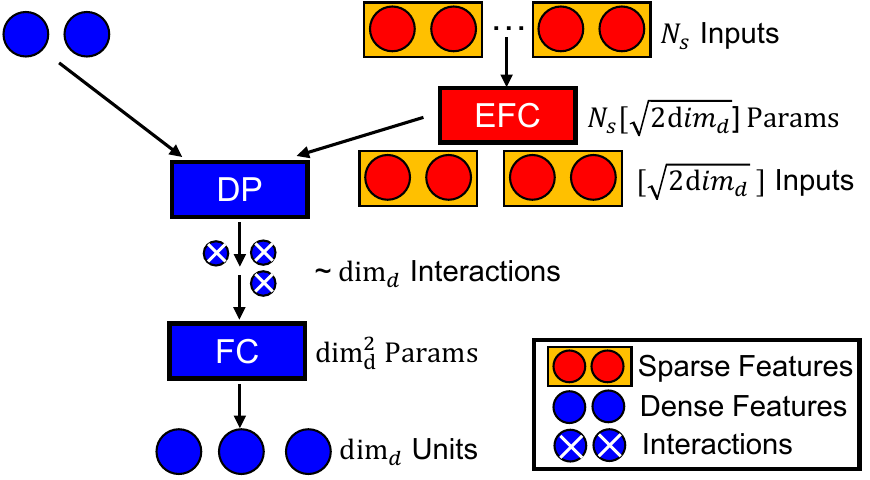}
    \caption{Operator-balancing interaction ensures linear parameter consumption and balance building operators.}
    \label{fig:op_imbalance}
    \end{center}
\end{figure}

Given $N_s$ sparse embeddings, a Dot-Product block produces $N_s^2/2$ pairwise interactions as a quadratic function on the number of sparse embeddings.
As detailed in Section \ref{sec:search_space}, the supernet requires a linear projection layer (i.e., FC) to match the output dimensions of operators within each choice block.
Typically, this leads to an extra $(N_s^2 \cdot dim_{d} /2)$ trainable weights for Dot-Products.

However, the weight consumption of such a projection layer is large, given many sparse embeddings.
For example, given $N_s=448$ and $dim_{d}=512$ in a 7-block NASRec supernet, the projection layer induces over $50M$ parameters in the NASRec supernet, which has a similar scale of parameter consumption with sparse embedding layers.
Such tremendous weight parameterization is a quadratic function of the number of sparse inputs $N_s$, yet other building operators have much fewer weights. For example, the number of trainable weights in EFC is a linear function of the sparse inputs $N_s$.
As a result, the over-parameterization in Dot-Product leads to an increased convergence rate for the Dot-Product operator and consequently favors parameter-consuming subnets with a high concentration of Dot-Product operations, as we observed.
In addition, the ignorance of heterogeneous operators other than Dot-Product provides a poor ranking of subnets, leading to sub-optimal performance on recommender systems.

We insert a simple EFC as a projection layer before the Dot-Product to mitigate such over-parameterization demonstrated in Figure \ref{fig:op_imbalance}.
Our intuition is projecting the number of sparse embeddings in Dot-Product to $[\sqrt{2dim_{d}}]$, such that the following Dot-Product operator produces approximately $dim_{d}$ outputs that later require a minimal projection layer to match the dimension.
As such, the Dot-Product operator consumes at most $(dim_{d}^2+N_s[\sqrt{2dim_{d}}])$ trainable weights and ensures a linear growth of parameter consumption with the number of sparse EFC $N_s$.
Thus, we balance the interaction operators to allow a similar convergence rate for all building operators.
We evaluate the training efficiency and ranking quality for supernets trained with/without operator-balancing interaction. Results demonstrate that operator-balancing interaction achieves 0.11 Kendall's $\tau$ improvement while reducing the search cost from 4 GPU hours to only 1.5 GPU hours.

\begin{table}[b]
\begin{center}
\caption{Effects of post-training fine-tuning on different path sampling strategies on \textit{NASRec-Full}. We demonstrate Pearson's $\rho$ and Kendall's $\tau$ over 100 random subnets on Criteo.}
\scalebox{1.0}{
    \begin{tabular}{|c|cc|cc|}
    \hline
    \multirow{2}{*}{\textbf{Path Sampling Strategy}} & \multicolumn{2}{|c|}{\textbf{No Fine-tuning}} &
    \multicolumn{2}{|c|}{\textbf{Fine-tuning}}  \\
    & Pearson's $\rho$  & Kendall's $\tau$ & Pearson's $\rho$  & Kendall's $\tau$ \\
    \hline
    Any-operator Any-connection & \large 0.37 & \large 0.28 & \large  0.46 & \large  0.43 \\
    Single-operator Single-connection & \large  0.05 & \large  0.02 & \large  0.43 & \large  0.29 \\
    Single-operator Any-connection & \large  0.46 & \large  \large  0.43 & \large  0.57 & \large  0.43 \\
    \hline
    \end{tabular}
}
    \label{tab:ft_subnets}    
\end{center}
\end{table}

\label{sec:path_ft}
\subsection{Post-training Fine-tuning}
Although dropout-like subnet training can effectively reduce the adaptation of weights for a specific subnet during supernet training, it may fail when weights should not be shared across certain subnets, leading to inaccurate subnet performance predictions by the supernet. To address this issue, we propose a post-training fine-tuning technique that re-adapts the weights of each standalone subnet back to its specific configuration after supernet training. This helps to re-calibrate the corrupted weights during supernet training while training other subnets.
In practice, fine-tuning only the last fully connected layer on the target dataset for a few training steps (e.g., 0.5K) is sufficient. This novel post-training fine-tuning technique comes with only marginal additional search cost and significantly boosts the ranking of subnets by addressing the underlying weight adaptation issue. As a result, this technique provides a better chance to discover better models for recommender systems.

Table \ref{tab:ft_subnets} demonstrates the improvement of post-training fine-tuning on different path sampling strategies.
Surprisingly, post-training fine-tuning achieves decent ranking quality improvement under Single-operator Single-connection and Any-operator Any-connection path sampling strategy.
This is because subnets under these strategies do not usually converge well in the supernet: they either suffer from poor supernet coverage or poor convergence induced by co-adaptation.
The fine-tuning process releases their potential and approaches their real performance on the target dataset.
Remarkably, the Single-operator Any-connection path sampling strategy cooperates well with post-training fine-tuning and achieves the global optimal Pearson's $\rho$ and Kendall's $\tau$ ranking correlation among different approaches, with at least $0.14$  Pearson's $\rho$ and Kendall's $\tau$ improvement on \textit{NASRec-Full} search space over Single-operator Single-connection sampling with fine-tuning.

\subsection{Evolutionary Search on Best Models}
We utilize regularized evolution~\cite{real2019regularized} to obtain the best child subnet in NASRec search space, including \textit{NASRec Small} and \textit{NASRec-Full}. Here, we first introduce a single mutation of a hierarchical genotype with the following sequence of actions in one of the choice blocks:
\begin{itemize}[noitemsep,leftmargin=*]
\item Re-sample the dimension of one dense building operator.
\item Re-sample the dimension of one sparse building operator.
\item Re-sample one dense building operator.
\item Re-sample one sparse building operator.
\item Re-sample its connection to other choice blocks.
\item Re-sample the choice of dense-to-sparse/sparse-to-dense merger that enables the communication between dense/sparse outputs.
\end{itemize}

%% file: _txt/5_Experiments.tex
\section{Experiments}
\label{sec:nasrec_exps}
We first show the detailed configuration that NASRec employs during the architecture search, model selection, and final evaluation.
Then, we demonstrate empirical evaluations on three popular recommender system benchmarks for Click-Through Rates (CTR) prediction: Criteo\footnote{\hyperlink{https://www.kaggle.com/c/criteo-display-ad-challenge}{https://www.kaggle.com/c/criteo-display-ad-challenge}}, Avazu\footnote{\hyperlink{https://www.kaggle.com/c/avazu-ctr-prediction/data}{https://www.kaggle.com/c/avazu-ctr-prediction/data}} and KDD Cup 2012\footnote{\hyperlink{https://www.kaggle.com/c/kddcup2012-track2/data}{https://www.kaggle.com/c/kddcup2012-track2/data}}. All three datasets are pre-processed in the same fashion as AutoCTR~\cite{song2020towards}. We release our implementation framework in \href{https://github.com/facebookresearch/NasRec}{NASRec}. On Criteo/Avazu/KDD Cup, we observe +/- 0.0002 as the standard deviation between each run and treat 0.001 as the level of significant improvement.

We show the statistics of each CTR benchmark in Table \ref{tab:dataset_ch}.
\begin{table}[h]
\begin{center}
    \caption{Statistics of different CTR benchmarks.}
    \begin{tabular}{|c|c|c|c|}
        \hline
         \textbf{Benchmark} & \# \textbf{Dense} & \# \textbf{Sparse} & \# \textbf{Samples (M)}  \\
         \hline
         Criteo & 13 & 26 & 45.84 \\
         Avazu & 0 & 23 & 40.42 \\
         KDD & 3 & 10 & 149.64  \\
         \hline
    \end{tabular}
    \label{tab:dataset_ch}    
\end{center}
\end{table}

Here, we observe that Criteo has the most dense (sparse) features and thus is the most complex and challenging benchmark. Avazu contains only dense features, thus requiring fewer interactions between dense outputs in each choice block. KDD has the least number of features and the most data, making it a relatively easier benchmark to train and evaluate.

\subsection{Search Configuration}
We first demonstrate the detailed configuration of \textit{NASRec-Full} search space as follows:\
\begin{itemize}[noitemsep,leftmargin=*]
    \item \textbf{Connection Search Components.} We utilize $N=7$ blocks in our NASRec search space. This allows a fair comparison with recent NAS methods~\cite{song2020towards}. All choice blocks can arbitrarily connect to previous choice blocks or raw features.
    \item \textbf{Operator Search Components.} In each choice block, our search space contains 6 distinct building operators, including 4 dense building operators: FC, Gating, Sum, and Dot-Product, and 2 distinct sparse building operators, EFC and Attention.
    The dense-sparse merger option is fully explored. 
    \item \textbf{Dimension Search Components.} For each dense building operator, the dense output dimension can be chosen from \{16, 32, 64, 128, 256, 512, 768, 1024\}. The sparse output dimension can be chosen from \{16, 32, 48, 64\} for each sparse building operator.
    \item \textbf{Quantization Search Components.} For each dense/sparse building operator, we perform weight/activation quantization of 4/8 bits for each building operator. This provides 16384 extra search complexity for a $N=7$ block search space.
\end{itemize}
In \textit{NASRec-Small}, we employ the same settings except that we use only 2 dense building operators: FC, Dot-Product, and 1 sparse building operator: EFC. 
Then, we illustrate some techniques for brewing the NASRec supernet, including the configuration of embedding, supernet warm-up, and supernet training settings.
\begin{itemize}[noitemsep,leftmargin=*]
    \item \textbf{Capped Embedding Table.} 
    We cap the maximum embedding table size to 0.5M during supernet training for search efficiency.
    During the final evaluation, we maintain the full embedding table to retrieve the best performance, i.e., 540M parameters in DLRM~\cite{naumov2019deep} on Criteo to ensure a fair comparison.
    
    \item \textbf{Supernet Warm-up.}
    The supernet may collapse at initial training phases due to the varying sampled paths and uninitialized embedding layers.
    To mitigate the supernet's initial collapse, we randomly sample the full supernet at the initial $1/5$ of the training steps, with a probability $p$ that linearly decays from 1 to 0. 
    This provides dimension warm-up, operator warm-up~\cite{bender2020can}, and connection warm-up for the supernet with minimal impact on the quality of sampled paths.
    
    \item \textbf{Supernet Training Settings.} 
    We insert layer normalization~\cite{ba2016layer} into each building operator to stabilize supernet training.
    Our choice of hyperparameters is robust over different NASRec search spaces and recommender system benchmarks. 
    We train the supernet for only one epoch with Adagrad optimizer, an initial learning rate of 0.12, and a cosine learning rate schedule~\cite{loshchilov2016sgdr} on target recommender system benchmarks. 
\end{itemize}

Finally, we present the details of regularized evolution and model selection strategies over NASRec search spaces.
\begin{itemize}[noitemsep,leftmargin=*]
\item \textbf{Regularized Evolution.} Despite the large size of \textit{NASRec-Full} and \textit{NASRec-small}, we employ an efficient configuration of regularized evolution to seek the optimal subnets from the supernet.
Specifically, we maintain a population of 128 architectures and run regularized evolution for 240 iterations. In each iteration, we first pick up the best architecture from 64 sampled architectures from the population as the parent architecture and generate 8 child architectures to update the population.

\item \textbf{Model Selection.} We follow the evaluation protocols in AutoCTR~\cite{song2020towards} and split each target dataset into 3 sets: training (80\%), validation (10\%), and testing (10\%). During the weight-sharing neural architecture search, we train the supernet on the training set and select the top 15 subnets on the validation set.
We train the top 15 models from scratch and select the best subnet, NASRecNet, as the final architecture. We perform light tuning on the learning rate of the best subnet within range (0.1, 0.2) and demonstrate the best learning rate setting on the open-source repository\footnote{https://github.com/facebookresearch/NasRec}.
\end{itemize}

\begin{table*}[t]
    \begin{center}
    \caption{Performance of NASRec on General CTR Predictions Tasks.}
    \scalebox{0.85}{
    \begin{tabular}{|c|c|cc|cc|cc|c|c|}
    \hline
     & \multirow{2}{*}{\textbf{Method}} & \multicolumn{2}{|c|}{\textbf{Criteo}}  &  \multicolumn{2}{|c|}{\textbf{Avazu}} & \multicolumn{2}{|c|}{\textbf{KDD Cup 2012}}  & \textbf{Search Cost} \\
      & &  Log Loss & AUC & Log Loss & AUC & Log Loss & AUC & (GPU days) \\
    \hline \hline
    \multirow{4}{*}{\textbf{Hand-crafted Arts}} & DLRM~\cite{naumov2019deep} & 0.4436 & 0.8085 & 0.3814 & 0.7766 & 0.1523 & 0.8004 & - \\
    & xDeepFM~\cite{lian2018xdeepfm} & 0.4418 & 0.8052 & - & - & - & - & - \\
    & AutoInt+~\cite{song2019autoint} & 0.4427 & 0.8090 & 0.3813 & 0.7772 & 0.1523 & 0.8002 & - \\
    & DeepFM~\cite{guo2017deepfm} & 0.4432 & 0.8086 & 0.3816 & 0.7767 & 0.1529 & 0.7974 & -\\
    \hline

    \multirow{7}{*}{\textbf{NAS-crafted Arts}} & DNAS~\cite{krishna2021differentiable} & 0.4442 & - & - & - & - & - & - \\
    & PROFIT~\cite{gao2021progressive} & 0.4427 & 0.8095 & \textbf{0.3735} & 0.7883 & - & - & $\sim$0.5 \\
    & AutoCTR~\cite{song2020towards} & 0.4413 & 0.8104 & 0.3800 & 0.7791 & 0.1520 & 0.8011 & $\sim$0.75 \\
    & Random Search @ \textit{NASRec-Small} & 0.4411 & 0.8105 & 0.3748 & 0.7885 & 0.1500 & 0.8123 & 1.0 \\
    & Random Search @ \textit{NASRec-Full} & 0.4418 & 0.8098 & 0.3767 & 0.7853 & 0.1509 & 0.8071 & 1.0 \\
    & AutoML @ \textit{NASRec-Small} & \textbf{0.4399} &\textbf{0.8118} & 0.3747 & 0.7887 & \textbf{0.1495} & \textbf{0.8135} & $\sim$0.25 \\
    & AutoML @ \textit{NASRec-Full} & \textbf{0.4408} & \textbf{0.8107} & \textbf{0.3737} & \textbf{0.7903} & \textbf{0.1491} & \textbf{0.8154} & $\sim$0.3 \\
    \hline
    \end{tabular}
    \label{tab:ctr_results}
    }
    \end{center}
\end{table*}

\subsection{Recommender System Benchmark Results}
We train our AutoML-crafted models from scratch on three classic recommender system benchmarks and compare the performance of models that NASRec crafts on three general recommender system benchmarks. In Table~\ref{tab:ctr_results}, we report the evaluation results of our end-to-end crafted models and a random search baseline, which randomly samples and trains models in our NASRec search space.

\noindent \textbf{State-of-the-art Performance.}
Even within an aggressively large \textit{NASRec-Full} search space, our crafted models achieve record-breaking performance over hand-crafted CTR models~\cite{naumov2019deep,guo2017deepfm,lian2018xdeepfm} with minimal human priors as shown in Table \ref{tab:ctr_results}.
Compared with AutoInt~\cite{song2019autoint}, the hand-crafted model that fabricates feature interactions with delicate engineering efforts, our crafted model achieves $\sim$0.003 Log Loss reduction on Criteo, $\sim$0.007 Log Loss reduction on Avazu, and $\sim$0.003 Log Loss reduction on KDD Cup 2012, with minimal human expertise and interventions.

Next, we compare our crafted models to the more recent NAS-crafted models. 
Compared to AutoCTR~\cite{song2020towards}, NASRecNet achieves the state-of-the-art (SOTA) Log Loss, and AUC on all three recommender system benchmarks.
With the same scale of search space as AutoCTR (i.e., \textit{NASRec-Small} search space), our crafted model yields 0.001 Log Loss reduction on Criteo, 0.005 Log Loss reduction on Avazu, and 0.003 Log Loss reduction on KDD Cup 2012.
Compared to DNAS~\cite{krishna2021differentiable} and PROFIT~\cite{gao2021progressive}, which only focuses on configuring part of the architectures, such as dense connectivity, our crafted model achieves at least $\sim$ 0.002 Log Loss reduction on Criteo, justifying the significance of full architecture search on recommender systems.

By extending NASRec to an extremely large \textit{NASRec-Full} search space, our crafted model further improves its result on Avazu and outperforms PROFIT by $\sim$ 0.002 AUC improvement with on-par Log Loss, justifying the design of \textit{NASRec-Full} with aggressively large cardinality and minimal human priors. On Criteo and KDD Cup 2012, NASRec maintains the edge in discovering state-of-the-art CTR models compared to existing NAS methods~\cite{song2020towards,gao2021progressive,krishna2021differentiable}.

\noindent \textbf{Efficient Search within a Versatile Search Space.}
Despite a larger NASRec search space that presents more challenges to fully explore, NASRec achieves at least 1.7$\times$ searching efficiency compared to state-of-the-art efficient NAS methods~\cite{song2020towards,gao2021progressive} with significant Log Loss improvement on all three benchmarks.
This is greatly attributed to the efficiency of Weight-Sharing NAS on heterogeneous operators and multi-modality data.

We observe that a compact \textit{NASRec-Small} search space produces strong random search baselines, while a larger \textit{NASRec-Full} search space has a weaker baseline.
A limited search budget makes it more challenging to discover promising models within a large search space.
Yet, the scalable WS-NAS tackles the exploration of full \textit{NASRec-Full} search space thanks to the broad coverage of the supernet. With an effective Single-Operator Any-connection path sampling strategy, WS-NAS improves the quality of discovered models on Criteo and discovers a better model on Avazu and KDD Cup 2012 than the NASRec-Small search space.

\noindent \textbf{Co-design Evaluation.}
Following the aforementioned search procedures on NASRec search space, we further enable quantization design space and inherit the same configurations, including dense/sparse building operator choices, hyperparameters, and regularized evolution configurations.
Instead of searching on NASRec-Full search space, we use NASRec-Small search space as all of the included building operators are PIM-compatible.
We model buffers using CACTI~\cite{cacti} at 32nm. We use the same ReRAM parameters as modeled in MNSIM2.0~\cite{mnsim} to obtain the area, latency, and power consumption parameters of the ReRAM crossbars. We build an in-house simulator to simulate the performance. The co-exploration process and performance simulation are performed in Intel Xeon Gold 6254 processors. We use NVIDIA A5000 devices to speed up the co-exploration.
We perform quantitative simulation using Criteo dataset features, i.e., 13 integer dense features and 26 categorical sparse features.
Our simulation results demonstrate a 1.5$\times$ speedup for the empirically handcrafted ReRAM design and a 1.1$\times$ speedup for RecNMP~\cite{ke2019recnmp} under the NASRec-Small search space. Additionally, the searched design shows 1.8$\times$ and 5.2$\times$ energy efficiency compared to the empirical design and RecNMP.
We will use the investigation and discovery of the NASRec-Full search space in future work.

\begin{table}[t]
    \caption{Model Complexity Analysis.}
    \begin{center}
    \scalebox{1.0}{
    \begin{tabular}{|c|ccc|ccc|}
    \hline
         \multirow{2}{*}{\textbf{Method}} & \multicolumn{3}{|c|}{\textbf{Log Loss}} & \multicolumn{3}{|c|}{\textbf{FLOPS(M)}}  \\   
         & Criteo & Avazu & KDD & Criteo & Avazu & KDD \\
         \hline
         DLRM & 0.4436 & 0.3814 & 0.1523 & 26.92 & 18.29 & 25.84 \\
         DeepFM & 0.4432 & 0.3816 & 0.1529 & 22.74 & 22.50 & 21.66 \\
         AutoInt+ & 0.4427 & 0.3813 & 0.1523 & 18.33 & 17.49 & 14.88 \\
    \hline
    AutoCTR & 0.4413 & 0.3800 & 0.1520 & 12.31 & 7.12 & 3.02 \\
    AutoML @ \textit{NASRec-Small} & \textbf{0.4399} & 0.3747 & 0.1495 & 2.20 & 3.08 & 3.48 \\
    AutoML @ \textit{NASRec-Full} & 0.4408 & \textbf{0.3737} & \textbf{0.1491} & \textbf{1.45} & \textbf{1.87} & \textbf{1.09} \\
    \hline
    \end{tabular}        
    }
    \end{center}
    \label{tab:model_complexity}
\end{table}

\subsection{Discussion}
In this section, we analyze the complexity of our crafted models and demonstrate the impact of our proposed techniques for mitigating ranking disorders and improving the quality of searched models.

\noindent \textbf{Model Complexity Analysis.}
We compare the complexity of our crafted models with that of SOTA hand-crafted and NAS models.
We collect all baselines from AutoCTR~\cite{song2020towards} and compare performance versus the number of Floating-point Operations (FLOPs) in Table \ref{tab:model_complexity}. 

We profile all FLOPS of our crafted models using FvCore~\cite{fvcore}.
Even without any FLOPs constraints, our crafted models outperform existing models efficiently. Despite achieving lower Log Loss, our crafted models reduce FLOPS by 8.5$times$, 3.8$times$, and 2.8$times$ on Criteo, Avazu, and KDD Cup 2012 benchmarks.
One possible reason is using operator-balancing interaction modules, which project the sparse inputs to a smaller dimension before carrying out cross-term feature interaction. This leads to significantly lower computation costs, contributing to compact yet high-performing recommender models.

\noindent \textbf{Effects of Path Sampling \& Fine-tuning.} We discuss the path sampling and fine-tuning techniques in Section \ref{sec:path_ft} and demonstrate the empirical evaluation of these techniques on the quality of searched models in Table \ref{tab:effect_pt}. The results show that (1) the importance of path sampling far outweighs the importance of fine-tuning in deciding the quality of searched models, and (2) a higher Kendall's $\tau$ that correctly ranks subnets in NASRec search space (i.e., Table \ref{tab:effect_pt}) indicates a consistent improvement on searched models.

\begin{table}[t]
    \caption{Effects of different training techniques on NASRecNet, evaluated on Criteo.}
    \begin{center}
    \scalebox{1.0}{

    \begin{tabular}{|c|c|c|}
    \hline
        \textbf{Method} & \textbf{Log Loss} & \textbf{FLOPS(M)}  \\
        \hline
        Baseline (Single-operator Any-connection + Fine-tuning) & \large 0.4408 & \large 1.45  \\
        Single-operator Single-connection + Fine-tuning & \large 0.4417 & \large 1.78 \\
        Any-operator Any-connection + Fine-tuning & \large 0.4413 & \large \large 2.04 \\
        Single-operator Any-connection, NO Fine-tuning &  \large 0.4410 & \large 3.62 \\
        \hline
    \end{tabular}
    }
    \end{center}
    \label{tab:effect_pt}
\end{table}

%% file: _txt/7_Supplementary_Material.tex
\section{Ablation Studies}
\label{sec:nasrec_ablation}

In this section, we provide more details regarding NASRec, including (1) the visualization and insight of searched architectures, (2) an ad-hoc structured pruning of AutoML-crafted models for enhanced model efficiency on Criteo/Avazu, and (3) the details on subnet sampling and ranking. 

\subsection{Model Visualization}
We visualize the models searched within the NASRec-Small/NASRec-Full search space on three CTR benchmarks: Criteo, Avazu, and KDD.

\begin{figure}[h]
\vspace{-1em}
\begin{center}
    \includegraphics[width=0.8\linewidth]{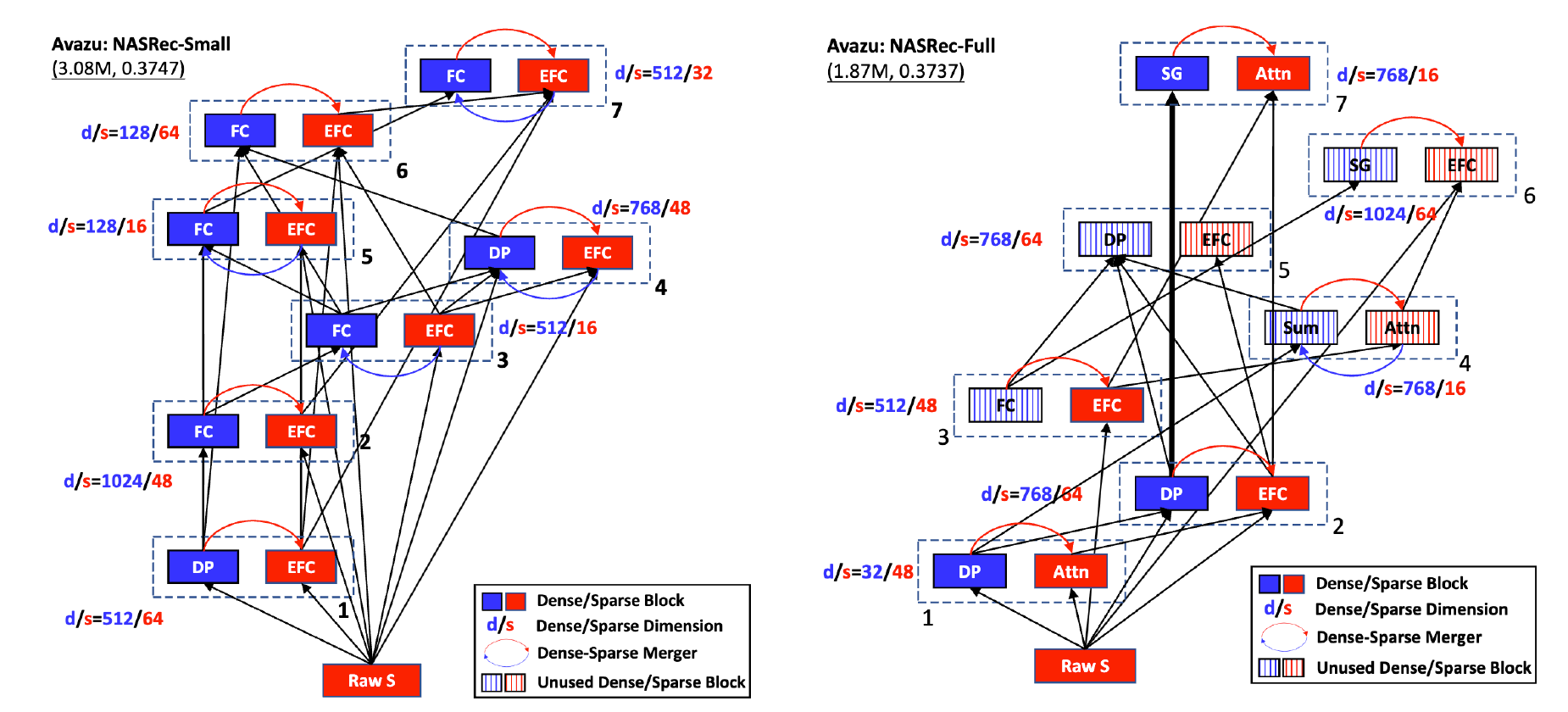}
    \caption{Best NASRec models discovered on Avazu.}  
    \vspace{-1em}
    \label{fig:avazu_best_nasrec}    
\end{center}
\end{figure}

\noindent \textbf{Avazu.} Figure \ref{fig:avazu_best_nasrec} depicts the detailed structures of the best architecture within the NASRec-Small/NASRec-Full search space. Here, a striped blue (red) block indicates an unused dense (sparse) block in the final architecture, and a bold connection indicates the same source input for a dense operator with two inputs (i.e., Sigmoid Gating and Sum).

As the Avazu benchmark only contains sparse features, the interaction and extraction of dense representations are less important. For example, the best model within NASRec-Full search space only contains one operator (i.e., Sigmoid Gating) that solely processes dense representations, yet with more Dot-Product (DP) and Attention (Attn) blocks that interact with the sparse representations.
Within the NASRec-Small search space, FC layers process dense representations more frequently after interacting with the sparse representations in the Dot-Product block. Yet, processing dense features requires slightly more fully connected blocks than the self-attention mechanism adopted in the NASRec-Full search space.

\begin{figure}[h]  
    \begin{center}
        \includegraphics[width=0.8\linewidth]{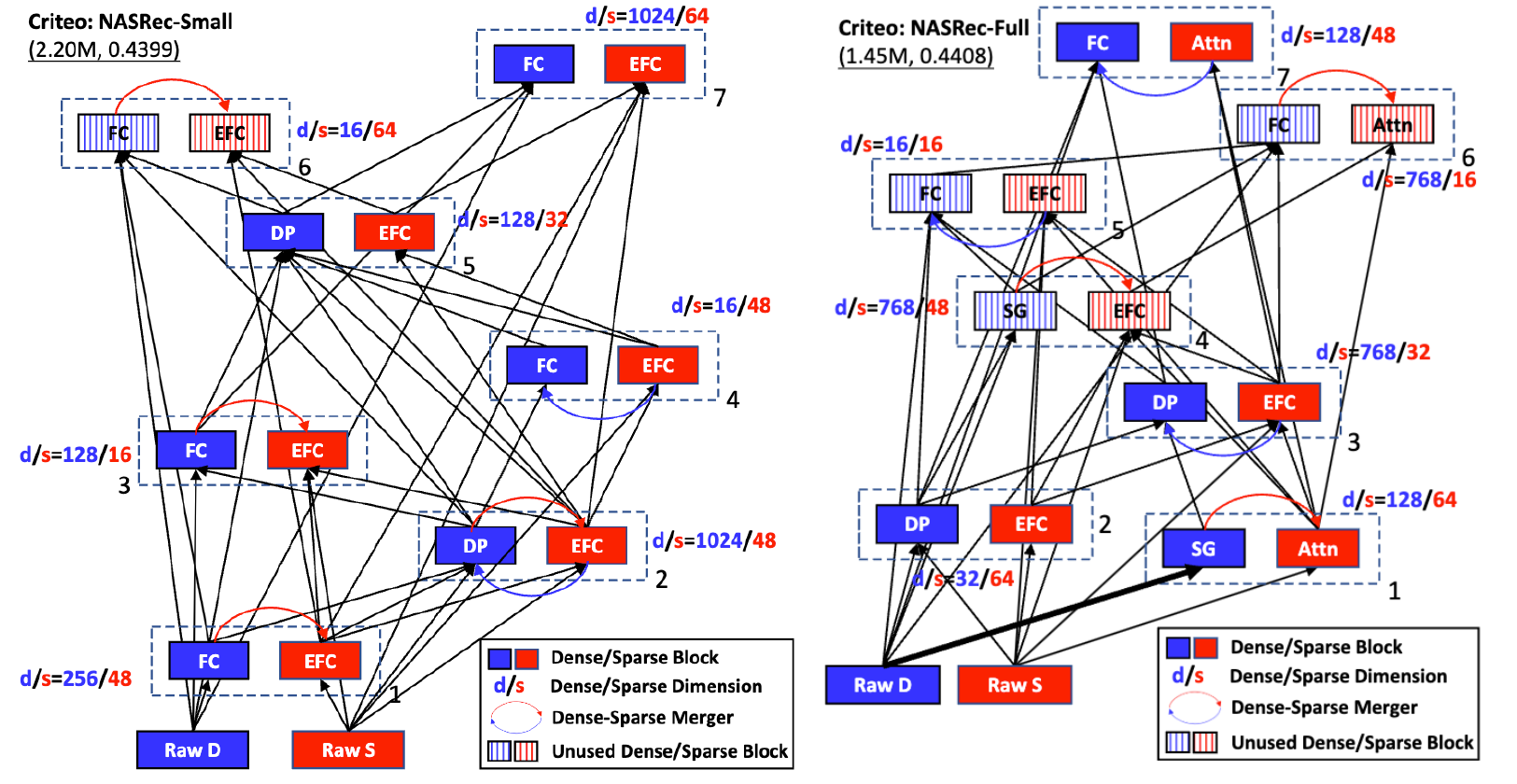}
        \caption{Best NASRec models discovered on Criteo.}   
        \vspace{-1em}
        \label{fig:criteo_best_nasrec}
    \end{center}
\end{figure}

\noindent \textbf{Criteo.} Figure \ref{fig:criteo_best_nasrec} depicts the detailed structures of the best architecture within the NASRec-Small/NASRec-Full search space. Here, a striped blue (red) block indicates an unused dense (sparse) block in the final architecture, and a bold connection indicates the same source input for a dense operator with two inputs (i.e., Sigmoid Gating and Sum).

\begin{figure}[b]
    \begin{center}
        \includegraphics[width=0.8\linewidth]{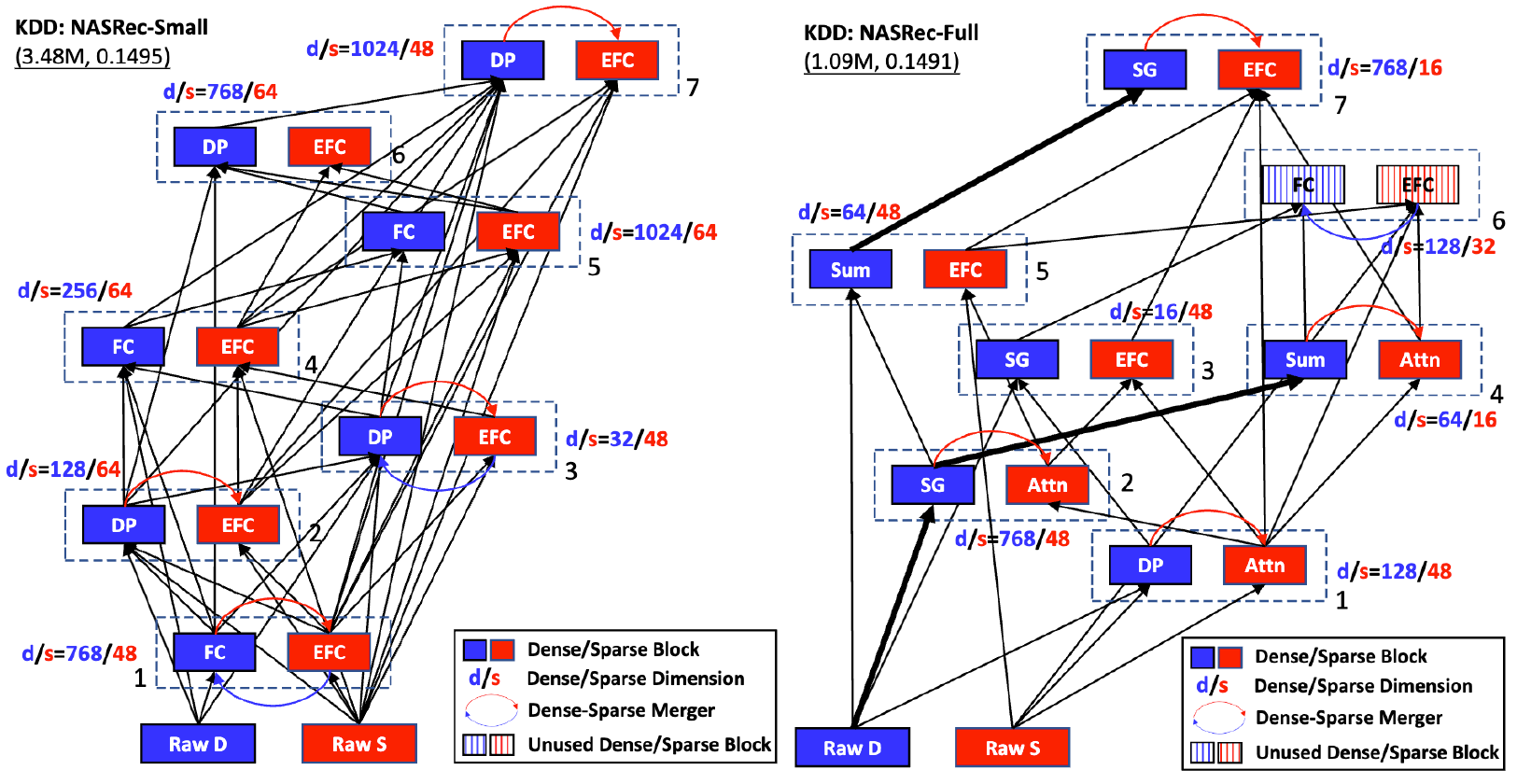}
        \caption{Best NASRec models discovered on KDD.}
        \vspace{-1em}
        \label{fig:kdd_best_nasrec}        
    \end{center}
\end{figure}

Criteo contains the richest set of dense (sparse) features and, thus, is the most complex in architectural fabrication. We observe that dense connectivity is highly appreciated within both NASRec-Small and NASRec-Full search space, indicating that feature fusion significantly impacts the log loss on complex benchmarks. In addition, self-gating on raw, dense features (i.e., block one @ NASRec-Full) is considered an important motif in interacting features. Similar patterns can also be observed in the best architecture searched on KDD benchmarks.

Due to the complexity of Criteo and NASRec-Full search blocks, the best-searched architecture does not use all seven blocks in the search space. Some of the blocks are not utilized in the final architecture. For example, the best architecture searched within NASRec-Full contains only four valid blocks. We leave this as a future work to improve supernet training so that deeper architectures can be discovered in a more scalable fashion.

\noindent \textbf{KDD.} Figure \ref{fig:kdd_best_nasrec} depicts the detailed structures of the best architecture within the NASRec-Small/NASRec-Full search space. Here, a striped blue (red) block indicates an unused dense (sparse) block in the final architecture, and a bold connection indicates the same source input for a dense operator with two inputs (i.e., Sigmoid Gating and Sum). Similar to what we found on Criteo, the searched architecture within NASRec-Full has more building operators yet less dense connectivity.

As KDD is a simpler benchmark with fewer dense (sparse) features, the architecture searched is simpler, especially within the NASRec search space. Similar self-gating on dense inputs is still important in designing a better architecture.

In the end, we summarize our observations on three unique benchmarks as follows:
\begin{itemize}[noitemsep,leftmargin=*]
    \item \textbf{Benchmark Complexity Decides Architecture Complexity.} The choice of a benchmark decides the complexity of the final architecture. The more complex a benchmark is, the more complicated a searched model is in dense connectivity and operator heterogeneity.

    \item \textbf{Search Space Decides Connectivity.} The best architecture searched within NASRec-Full on all three CTR benchmarks contains more operator heterogeneity and less dense connectivity. Yet, the reduced dense connectivity between different choice blocks helps reduce FLOPs consumption of searched models, leading to less complexity and better model efficiency. This also shows that the search for building operators may outweigh the importance of the search for dense connectivity when crafting an efficient CTR model.

    \item \textbf{Attention Has a Huge Impact.} Attention blocks are rarely studied in the existing literature on recommender systems. The architectures searched on NASRec-Full search space justify the effectiveness of the attention mechanism on aggregating dense (sparse) features. For example, the first block in the best architecture always adopts an attention layer to interact \textit{raw, sparse inputs}. The stacking of attention blocks is also observed in searched architectures to demonstrate high-order interaction between dense (sparse) features.

    \item \textbf{Self-Gating Is a Useful Motif.} Self-gating indicates a pairwise gating operator with identical dense inputs. On both Criteo/KDD benchmarks, self-gating is discovered to process \textit{raw, dense inputs} and provide higher-quality dense projections. On Avazu, with no dense input features, self-gating is discovered to combine a higher-level dense representation for better prediction results.
\end{itemize}

\subsection{Pruning NASRec via Lottery Ticket}
Recommender systems face unique challenges due to the heterogeneity, uncertainty, and multi-modality of data. 
It is challenging to apply existing pruning techniques~\cite{kowald2017temporal, carterette2007evaluating, guo2015trustsvd} and maintain performance on compressed recommender models.
For example, existing pruning methods require the training of recommender models for several passes, leading to severe performance degradation and instability~\cite{zhou2018deep} due to overfitting.
Our methodology is inspired by the Lottery Ticket Hypothesis~\cite{frankle2018lottery} that learns a smaller sub-architecture (i.e., winning tickets) without involving multi-pass training.
Our methodology includes \textbf{mask generation} and \textbf{structured pruning}.

\noindent \textbf{Mask Generation.}
We design a mask generation process to mask out zero weights in the original weight matrix $W_{orig}$.
We generate a mask matrix $M$ using a 2-layer MLP for each weight matrix. 
The 2-layer MLP inputs the original weight matrix $W_{orig}$.
The first MLP layer employs a ReLU activation function, and the second MLP layer uses a sigmoid activation. 
The formulation is as follows:
\begin{equation}
    M = \sigma(W_2 \cdot \text{ReLU}(W_1 \cdot W_{\text{orig}})),
\end{equation}
where \(W_{\text{orig}}\) is the original weight matrix, \(W_1\)/\(W_2\) denotes the weights of the first/second layers of the MLP, and $\cdot$ denotes matrix multiplication. 
The first layer projects the weight matrix to a higher dimensional space to extract a rich representation. The second layer projects this high latent dimension back to the original dimensionality of the weight matrix. The final mask \(M\) is obtained through element-wise multiplication with the weight matrix:
\begin{equation}
    W_{\text{masked}} = M \odot W_{\text{orig}}.
\end{equation}
Here, $\odot$ denotes element-wise multiplication.

\noindent \textbf{Structured Pruning.}
We conduct iterative structured pruning by applying lottery tickets on recommender models to generate masks $M$ and zero out unmasked weights.
We initialize the original weight mask as $M^{(0)}$, with all mask values set to 1.
The overall iterative structured pruning takes $T$ iterations.
Within each iteration $t$, we train a backbone model with the learned lottery ticket $M^{(t-1)}$ from scratch and zero out 20\% of the lowest values in $M^{(t-1)}$ to derive a new mask $M^{(t)}.$.

\noindent \textbf{Experimental Evaluation.}
We apply the structure above pruning methodology to the NASRec model searched within the NASRec search space, which contains various building operators.
We apply the pruning methodology on all building blocks covering FC/EFC/DP modules on dense/sparse building operators and dense-to-sparse/sparse-to-dense mergers. 
We use ``Model'' to represent AutoML models crafted under \textit{NASRec-Full} search space, with baseline results presented in Table \ref{tab:ctr_results}.
We showcase our evaluation of the Criteo/Avazu dataset in Table \ref{tab:exp_pruning}, demonstrating that our pruning method effectively reduces FLOPs without significant degradation in loss.
Specifically, our approach reduces 53\% / 46\% FLOPs on NASRec models on the Criteo/Avazu benchmark without incurring noticeable log loss.
In some cases, combining lottery tickets with recommender models shows some gains (e.g., ``Model'' on Criteo), indicating potential model redundancy in existing searched models and possible headroom for improvement.

\begin{table*}[t]
    \begin{center}
    \caption{Pruning ``Model'' on general CTR Prediction Tasks.}
    \scalebox{0.97}{
    \begin{tabular}{|c|c|cc|cc|}
    \hline
    \multirow{2}{*}{} & \multirow{2}{*}{\textbf{Model}} & \multicolumn{2}{|c|}{\textbf{Mask-Based Pruning}} & \multicolumn{2}{|c|}{\textbf{Magnitude-Based Pruning}} \\
      & &  Log Loss $\downarrow$ & MFLOPs (Percentage) & Log Loss $\downarrow$ & MFLOPs (Percentage) \\
    \hline
    \multirow{4}{*}{\textbf{Criteo}} & Model & 0.4408 & 1.45 (100\%) & 0.4408 & 1.45 (100\%)  \\
     & Model+Pruning (T=5) &  0.4402 & 0.78 (54\%) & 0.4405 & 0.78 (54\%) \\
     & Model+Pruning (T=3) & 0.4403 & 1.01 (69\%) & 0.4406 & 1.01(69\%) \\
     & Model+Pruning (T=1) &0.4402 & 1.37 (94\%) & 0.4406 & 1.37 (94\%) \\
    \hline
    \multirow{4}{*}{\textbf{Avazu}} & Model & 0.3737 & 1.87 (100\%) & 0.3737 & 1.87 (100\%)\\
     & Model+Pruning (T=5) & 0.3742 & 0.88 (47\%) & 0.3748 & 0.88 (47\%) \\
     & Model+Pruning (T=3) & 0.3741 & 1.23 (66\%) & 0.3746 & 1.23 (66\%) \\
     & Model+Pruning (T=1) & 0.3741 & 1.58 (84\%) & 0.3744 & 1.58 (84\%) \\
    \hline
    \end{tabular}
    \label{tab:exp_pruning}
    }
    \end{center}
\end{table*}

\begin{figure}[t]
    \includegraphics[width=0.75\linewidth]{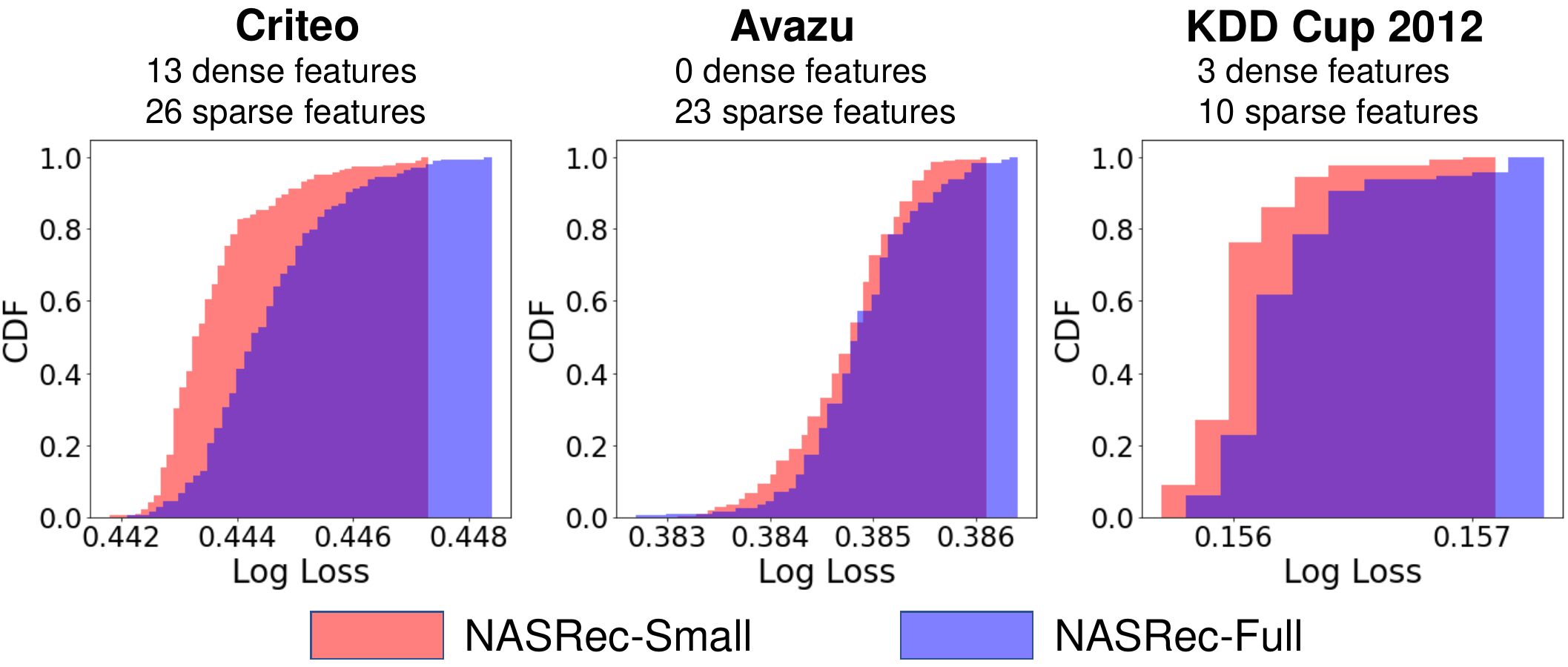}
    \caption{CDF of log loss on CTR benchmarks.}    
    \label{fig:subnet_cdf}
\end{figure}

\subsection{Subnet Sampling Details}
In Section 4, we sample 100 subnets within NASRec-Full search space on Criteo benchmark, with a more balanced and efficient setting on dimension search components: the dense output dimension can choose from \{32, 64, 128, 256, 512\}, and the sparse output dimension can choose from \{16, 32, 64\}.
All subnets are trained on the Criteo benchmark with a batch size of 1024 and a learning rate of 0.12.

We plot the Cumulative Distribution Function (CDF) of sampled subnets on all three benchmarks in Table \ref{fig:subnet_cdf}. For the top 50\% architectures evaluated on NASRec-Full supernet, we report a Kendall's $\tau$ of 0.24 for the Criteo benchmark, showing a clear improvement in ranking top-performing architectures over the random search (0.0). In future work, we propose establishing a CTR benchmark for NAS to increase the statistical significance of evaluated ranking coefficients and better facilitate the research in accurately ranking different architectures. 

%% file: _txt/8_Conclusion.tex
\section{Conclusion}
\label{sec:conclusion}
In this paper, we introduce a novel paradigm for fully enabling Automated
Machine Learning (AutoML) in full-stack recommender model design,
leveraging Weight Sharing Neural Architecture Search (WS-NAS) under
diverse data modalities and architectures. We construct a
large supernet that encompasses the entire architecture search space,
incorporating versatile building blocks and dense connection operators to
minimize human intervention in automated architecture design for
recommender systems.
To address the scalability and heterogeneity challenges inherent in
large-scale NASRec search spaces, we propose a series of techniques to
enhance training efficiency and mitigate ranking disorders. 
We achieve state-of-the-art performance on three prominent recommender system benchmarks, showcasing promising prospects for a full architecture search and motivating further research towards fully automated architecture
fabrication with minimal human priors.
Moreover, we suggest opportunities for co-designing models and
inference hardware and unlock the potential to perform ad-hoc structure pruning on AutoML-crafted models to achieve improved
performance.

%% file: sample-sigconf.bbl

\begin{thebibliography}{50}


\ifx \showCODEN    \undefined \def \showCODEN     #1{\unskip}     \fi
\ifx \showDOI      \undefined \def \showDOI       #1{#1}\fi
\ifx \showISBNx    \undefined \def \showISBNx     #1{\unskip}     \fi
\ifx \showISBNxiii \undefined \def \showISBNxiii  #1{\unskip}     \fi
\ifx \showISSN     \undefined \def \showISSN      #1{\unskip}     \fi
\ifx \showLCCN     \undefined \def \showLCCN      #1{\unskip}     \fi
\ifx \shownote     \undefined \def \shownote      #1{#1}          \fi
\ifx \showarticletitle \undefined \def \showarticletitle #1{#1}   \fi
\ifx \showURL      \undefined \def \showURL       {\relax}        \fi
\providecommand\bibfield[2]{#2}
\providecommand\bibinfo[2]{#2}
\providecommand\natexlab[1]{#1}
\providecommand\showeprint[2][]{arXiv:#2}

\bibitem[Ba et~al\mbox{.}(2016)]%
        {ba2016layer}
\bibfield{author}{\bibinfo{person}{Jimmy~Lei Ba}, \bibinfo{person}{Jamie~Ryan Kiros}, {and} \bibinfo{person}{Geoffrey~E Hinton}.} \bibinfo{year}{2016}\natexlab{}.
\newblock \showarticletitle{Layer normalization}.
\newblock \bibinfo{journal}{\emph{arXiv preprint arXiv:1607.06450}} (\bibinfo{year}{2016}).
\newblock


\bibitem[Balasubramonian et~al\mbox{.}(2017)]%
        {cacti}
\bibfield{author}{\bibinfo{person}{Rajeev Balasubramonian}, \bibinfo{person}{Andrew~B. Kahng}, \bibinfo{person}{Naveen Muralimanohar}, \bibinfo{person}{Ali Shafiee}, {and} \bibinfo{person}{Vaishnav Srinivas}.} \bibinfo{year}{2017}\natexlab{}.
\newblock \showarticletitle{{CACTI} 7: New Tools for Interconnect Exploration in Innovative Off-Chip Memories}.
\newblock \bibinfo{journal}{\emph{{ACM} Trans. Archit. Code Optim.}} \bibinfo{volume}{14}, \bibinfo{number}{2} (\bibinfo{year}{2017}), \bibinfo{pages}{14:1--14:25}.
\newblock


\bibitem[Bender et~al\mbox{.}(2018)]%
        {bender2018understanding}
\bibfield{author}{\bibinfo{person}{Gabriel Bender}, \bibinfo{person}{Pieter-Jan Kindermans}, \bibinfo{person}{Barret Zoph}, \bibinfo{person}{Vijay Vasudevan}, {and} \bibinfo{person}{Quoc Le}.} \bibinfo{year}{2018}\natexlab{}.
\newblock \showarticletitle{Understanding and simplifying one-shot architecture search}. In \bibinfo{booktitle}{\emph{International Conference on Machine Learning}}. PMLR, \bibinfo{pages}{550--559}.
\newblock


\bibitem[Bender et~al\mbox{.}(2020)]%
        {bender2020can}
\bibfield{author}{\bibinfo{person}{Gabriel Bender}, \bibinfo{person}{Hanxiao Liu}, \bibinfo{person}{Bo Chen}, \bibinfo{person}{Grace Chu}, \bibinfo{person}{Shuyang Cheng}, \bibinfo{person}{Pieter-Jan Kindermans}, {and} \bibinfo{person}{Quoc~V Le}.} \bibinfo{year}{2020}\natexlab{}.
\newblock \showarticletitle{Can weight sharing outperform random architecture search? an investigation with tunas}. In \bibinfo{booktitle}{\emph{Proceedings of the IEEE/CVF Conference on Computer Vision and Pattern Recognition}}. \bibinfo{pages}{14323--14332}.
\newblock


\bibitem[Cai et~al\mbox{.}(2019)]%
        {cai2019once}
\bibfield{author}{\bibinfo{person}{Han Cai}, \bibinfo{person}{Chuang Gan}, \bibinfo{person}{Tianzhe Wang}, \bibinfo{person}{Zhekai Zhang}, {and} \bibinfo{person}{Song Han}.} \bibinfo{year}{2019}\natexlab{}.
\newblock \showarticletitle{Once-for-all: Train one network and specialize it for efficient deployment}.
\newblock \bibinfo{journal}{\emph{arXiv preprint arXiv:1908.09791}} (\bibinfo{year}{2019}).
\newblock


\bibitem[Carterette and Jones(2007)]%
        {carterette2007evaluating}
\bibfield{author}{\bibinfo{person}{Ben Carterette} {and} \bibinfo{person}{Rosie Jones}.} \bibinfo{year}{2007}\natexlab{}.
\newblock \showarticletitle{Evaluating search engines by modeling the relationship between relevance and clicks}.
\newblock \bibinfo{journal}{\emph{Advances in neural information processing systems}}  \bibinfo{volume}{20} (\bibinfo{year}{2007}).
\newblock


\bibitem[Chen et~al\mbox{.}(2019)]%
        {chen2019behavior}
\bibfield{author}{\bibinfo{person}{Qiwei Chen}, \bibinfo{person}{Huan Zhao}, \bibinfo{person}{Wei Li}, \bibinfo{person}{Pipei Huang}, {and} \bibinfo{person}{Wenwu Ou}.} \bibinfo{year}{2019}\natexlab{}.
\newblock \showarticletitle{Behavior sequence transformer for e-commerce recommendation in alibaba}. In \bibinfo{booktitle}{\emph{Proceedings of the 1st International Workshop on Deep Learning Practice for High-Dimensional Sparse Data}}. \bibinfo{pages}{1--4}.
\newblock


\bibitem[Cheng et~al\mbox{.}(2016)]%
        {cheng2016wide}
\bibfield{author}{\bibinfo{person}{Heng-Tze Cheng}, \bibinfo{person}{Levent Koc}, \bibinfo{person}{Jeremiah Harmsen}, \bibinfo{person}{Tal Shaked}, \bibinfo{person}{Tushar Chandra}, \bibinfo{person}{Hrishi Aradhye}, \bibinfo{person}{Glen Anderson}, \bibinfo{person}{Greg Corrado}, \bibinfo{person}{Wei Chai}, \bibinfo{person}{Mustafa Ispir}, {et~al\mbox{.}}} \bibinfo{year}{2016}\natexlab{}.
\newblock \showarticletitle{Wide \& deep learning for recommender systems}. In \bibinfo{booktitle}{\emph{Proceedings of the 1st workshop on deep learning for recommender systems}}. \bibinfo{pages}{7--10}.
\newblock


\bibitem[Covington et~al\mbox{.}(2016)]%
        {covington2016deep}
\bibfield{author}{\bibinfo{person}{Paul Covington}, \bibinfo{person}{Jay Adams}, {and} \bibinfo{person}{Emre Sargin}.} \bibinfo{year}{2016}\natexlab{}.
\newblock \showarticletitle{Deep neural networks for youtube recommendations}. In \bibinfo{booktitle}{\emph{Proceedings of the 10th ACM conference on recommender systems}}. \bibinfo{pages}{191--198}.
\newblock


\bibitem[Deng et~al\mbox{.}(2021)]%
        {deng2021deeplight}
\bibfield{author}{\bibinfo{person}{Wei Deng}, \bibinfo{person}{Junwei Pan}, \bibinfo{person}{Tian Zhou}, \bibinfo{person}{Deguang Kong}, \bibinfo{person}{Aaron Flores}, {and} \bibinfo{person}{Guang Lin}.} \bibinfo{year}{2021}\natexlab{}.
\newblock \showarticletitle{DeepLight: Deep lightweight feature interactions for accelerating CTR predictions in ad serving}. In \bibinfo{booktitle}{\emph{Proceedings of the 14th ACM international conference on Web search and data mining}}. \bibinfo{pages}{922--930}.
\newblock


\bibitem[Dosovitskiy et~al\mbox{.}(2020)]%
        {dosovitskiy2020image}
\bibfield{author}{\bibinfo{person}{Alexey Dosovitskiy}, \bibinfo{person}{Lucas Beyer}, \bibinfo{person}{Alexander Kolesnikov}, \bibinfo{person}{Dirk Weissenborn}, \bibinfo{person}{Xiaohua Zhai}, \bibinfo{person}{Thomas Unterthiner}, \bibinfo{person}{Mostafa Dehghani}, \bibinfo{person}{Matthias Minderer}, \bibinfo{person}{Georg Heigold}, \bibinfo{person}{Sylvain Gelly}, {et~al\mbox{.}}} \bibinfo{year}{2020}\natexlab{}.
\newblock \showarticletitle{An image is worth 16x16 words: Transformers for image recognition at scale}.
\newblock \bibinfo{journal}{\emph{arXiv preprint arXiv:2010.11929}} (\bibinfo{year}{2020}).
\newblock


\bibitem[Frankle and Carbin(2018)]%
        {frankle2018lottery}
\bibfield{author}{\bibinfo{person}{Jonathan Frankle} {and} \bibinfo{person}{Michael Carbin}.} \bibinfo{year}{2018}\natexlab{}.
\newblock \showarticletitle{The lottery ticket hypothesis: Finding sparse, trainable neural networks}.
\newblock \bibinfo{journal}{\emph{arXiv preprint arXiv:1803.03635}} (\bibinfo{year}{2018}).
\newblock


\bibitem[Gao et~al\mbox{.}(2021)]%
        {gao2021progressive}
\bibfield{author}{\bibinfo{person}{Chen Gao}, \bibinfo{person}{Yinfeng Li}, \bibinfo{person}{Quanming Yao}, \bibinfo{person}{Depeng Jin}, {and} \bibinfo{person}{Yong Li}.} \bibinfo{year}{2021}\natexlab{}.
\newblock \showarticletitle{Progressive Feature Interaction Search for Deep Sparse Network}.
\newblock \bibinfo{journal}{\emph{Advances in Neural Information Processing Systems}}  \bibinfo{volume}{34} (\bibinfo{year}{2021}).
\newblock


\bibitem[Gao et~al\mbox{.}(2020)]%
        {gao2020modularized}
\bibfield{author}{\bibinfo{person}{Luyu Gao}, \bibinfo{person}{Zhuyun Dai}, {and} \bibinfo{person}{Jamie Callan}.} \bibinfo{year}{2020}\natexlab{}.
\newblock \showarticletitle{Modularized transfomer-based ranking framework}.
\newblock \bibinfo{journal}{\emph{arXiv preprint arXiv:2004.13313}} (\bibinfo{year}{2020}).
\newblock


\bibitem[Guo et~al\mbox{.}(2015)]%
        {guo2015trustsvd}
\bibfield{author}{\bibinfo{person}{Guibing Guo}, \bibinfo{person}{Jie Zhang}, {and} \bibinfo{person}{Neil Yorke-Smith}.} \bibinfo{year}{2015}\natexlab{}.
\newblock \showarticletitle{Trustsvd: Collaborative filtering with both the explicit and implicit influence of user trust and of item ratings}. In \bibinfo{booktitle}{\emph{Proceedings of the AAAI conference on artificial intelligence}}, Vol.~\bibinfo{volume}{29}.
\newblock


\bibitem[Guo et~al\mbox{.}(2017)]%
        {guo2017deepfm}
\bibfield{author}{\bibinfo{person}{Huifeng Guo}, \bibinfo{person}{Ruiming Tang}, \bibinfo{person}{Yunming Ye}, \bibinfo{person}{Zhenguo Li}, {and} \bibinfo{person}{Xiuqiang He}.} \bibinfo{year}{2017}\natexlab{}.
\newblock \showarticletitle{DeepFM: a factorization-machine based neural network for CTR prediction}.
\newblock \bibinfo{journal}{\emph{arXiv preprint arXiv:1703.04247}} (\bibinfo{year}{2017}).
\newblock


\bibitem[He et~al\mbox{.}(2014)]%
        {he2014practical}
\bibfield{author}{\bibinfo{person}{Xinran He}, \bibinfo{person}{Junfeng Pan}, \bibinfo{person}{Ou Jin}, \bibinfo{person}{Tianbing Xu}, \bibinfo{person}{Bo Liu}, \bibinfo{person}{Tao Xu}, \bibinfo{person}{Yanxin Shi}, \bibinfo{person}{Antoine Atallah}, \bibinfo{person}{Ralf Herbrich}, \bibinfo{person}{Stuart Bowers}, {et~al\mbox{.}}} \bibinfo{year}{2014}\natexlab{}.
\newblock \showarticletitle{Practical lessons from predicting clicks on ads at facebook}. In \bibinfo{booktitle}{\emph{Proceedings of the eighth international workshop on data mining for online advertising}}. \bibinfo{pages}{1--9}.
\newblock


\bibitem[Ke et~al\mbox{.}(2019)]%
        {ke2019recnmp}
\bibfield{author}{\bibinfo{person}{Liu Ke}, \bibinfo{person}{Udit Gupta}, \bibinfo{person}{Carole-Jean Wu}, \bibinfo{person}{Benjamin~Youngjae Cho}, \bibinfo{person}{Mark Hempstead}, \bibinfo{person}{Brandon Reagen}, \bibinfo{person}{Xuan Zhang}, \bibinfo{person}{David Brooks}, \bibinfo{person}{Vikas Chandra}, \bibinfo{person}{Utku Diril}, \bibinfo{person}{Amin Firoozshahian}, \bibinfo{person}{Kim Hazelwood}, \bibinfo{person}{Bill Jia}, \bibinfo{person}{Hsien-Hsin~S. Lee}, \bibinfo{person}{Meng Li}, \bibinfo{person}{Bert Maher}, \bibinfo{person}{Dheevatsa Mudigere}, \bibinfo{person}{Maxim Naumov}, \bibinfo{person}{Martin Schatz}, \bibinfo{person}{Mikhail Smelyanskiy}, {and} \bibinfo{person}{Xiaodong Wang}.} \bibinfo{year}{2019}\natexlab{}.
\newblock \bibinfo{title}{RecNMP: Accelerating Personalized Recommendation with Near-Memory Processing}.
\newblock
\newblock
\showeprint[arxiv]{1912.12953}~[cs.DC]


\bibitem[Kowald et~al\mbox{.}(2017)]%
        {kowald2017temporal}
\bibfield{author}{\bibinfo{person}{Dominik Kowald}, \bibinfo{person}{Subhash~Chandra Pujari}, {and} \bibinfo{person}{Elisabeth Lex}.} \bibinfo{year}{2017}\natexlab{}.
\newblock \showarticletitle{Temporal effects on hashtag reuse in twitter: A cognitive-inspired hashtag recommendation approach}. In \bibinfo{booktitle}{\emph{Proceedings of the 26th International Conference on World Wide Web}}. \bibinfo{pages}{1401--1410}.
\newblock


\bibitem[Krishna et~al\mbox{.}(2021)]%
        {krishna2021differentiable}
\bibfield{author}{\bibinfo{person}{Ravi Krishna}, \bibinfo{person}{Aravind Kalaiah}, \bibinfo{person}{Bichen Wu}, \bibinfo{person}{Maxim Naumov}, \bibinfo{person}{Dheevatsa Mudigere}, \bibinfo{person}{Misha Smelyanskiy}, {and} \bibinfo{person}{Kurt Keutzer}.} \bibinfo{year}{2021}\natexlab{}.
\newblock \showarticletitle{Differentiable NAS Framework and Application to Ads CTR Prediction}.
\newblock \bibinfo{journal}{\emph{arXiv preprint arXiv:2110.14812}} (\bibinfo{year}{2021}).
\newblock


\bibitem[Lian et~al\mbox{.}(2018)]%
        {lian2018xdeepfm}
\bibfield{author}{\bibinfo{person}{Jianxun Lian}, \bibinfo{person}{Xiaohuan Zhou}, \bibinfo{person}{Fuzheng Zhang}, \bibinfo{person}{Zhongxia Chen}, \bibinfo{person}{Xing Xie}, {and} \bibinfo{person}{Guangzhong Sun}.} \bibinfo{year}{2018}\natexlab{}.
\newblock \showarticletitle{xdeepfm: Combining explicit and implicit feature interactions for recommender systems}. In \bibinfo{booktitle}{\emph{Proceedings of the 24th ACM SIGKDD international conference on knowledge discovery \& data mining}}. \bibinfo{pages}{1754--1763}.
\newblock


\bibitem[Liang et~al\mbox{.}(2019)]%
        {liang2019darts+}
\bibfield{author}{\bibinfo{person}{Hanwen Liang}, \bibinfo{person}{Shifeng Zhang}, \bibinfo{person}{Jiacheng Sun}, \bibinfo{person}{Xingqiu He}, \bibinfo{person}{Weiran Huang}, \bibinfo{person}{Kechen Zhuang}, {and} \bibinfo{person}{Zhenguo Li}.} \bibinfo{year}{2019}\natexlab{}.
\newblock \showarticletitle{Darts+: Improved differentiable architecture search with early stopping}.
\newblock \bibinfo{journal}{\emph{arXiv preprint arXiv:1909.06035}} (\bibinfo{year}{2019}).
\newblock


\bibitem[Liu et~al\mbox{.}(2018)]%
        {liu2018darts}
\bibfield{author}{\bibinfo{person}{Hanxiao Liu}, \bibinfo{person}{Karen Simonyan}, {and} \bibinfo{person}{Yiming Yang}.} \bibinfo{year}{2018}\natexlab{}.
\newblock \showarticletitle{Darts: Differentiable architecture search}.
\newblock \bibinfo{journal}{\emph{arXiv preprint arXiv:1806.09055}} (\bibinfo{year}{2018}).
\newblock


\bibitem[Loshchilov and Hutter(2016)]%
        {loshchilov2016sgdr}
\bibfield{author}{\bibinfo{person}{Ilya Loshchilov} {and} \bibinfo{person}{Frank Hutter}.} \bibinfo{year}{2016}\natexlab{}.
\newblock \showarticletitle{Sgdr: Stochastic gradient descent with warm restarts}.
\newblock \bibinfo{journal}{\emph{arXiv preprint arXiv:1608.03983}} (\bibinfo{year}{2016}).
\newblock


\bibitem[Mellor et~al\mbox{.}(2021)]%
        {mellor2021neural}
\bibfield{author}{\bibinfo{person}{Joe Mellor}, \bibinfo{person}{Jack Turner}, \bibinfo{person}{Amos Storkey}, {and} \bibinfo{person}{Elliot~J Crowley}.} \bibinfo{year}{2021}\natexlab{}.
\newblock \showarticletitle{Neural architecture search without training}. In \bibinfo{booktitle}{\emph{International conference on machine learning}}. PMLR, \bibinfo{pages}{7588--7598}.
\newblock


\bibitem[Naumov et~al\mbox{.}(2019)]%
        {naumov2019deep}
\bibfield{author}{\bibinfo{person}{Maxim Naumov}, \bibinfo{person}{Dheevatsa Mudigere}, \bibinfo{person}{Hao-Jun~Michael Shi}, \bibinfo{person}{Jianyu Huang}, \bibinfo{person}{Narayanan Sundaraman}, \bibinfo{person}{Jongsoo Park}, \bibinfo{person}{Xiaodong Wang}, \bibinfo{person}{Udit Gupta}, \bibinfo{person}{Carole-Jean Wu}, \bibinfo{person}{Alisson~G Azzolini}, {et~al\mbox{.}}} \bibinfo{year}{2019}\natexlab{}.
\newblock \showarticletitle{Deep learning recommendation model for personalization and recommendation systems}.
\newblock \bibinfo{journal}{\emph{arXiv preprint arXiv:1906.00091}} (\bibinfo{year}{2019}).
\newblock


\bibitem[Real et~al\mbox{.}(2019)]%
        {real2019regularized}
\bibfield{author}{\bibinfo{person}{Esteban Real}, \bibinfo{person}{Alok Aggarwal}, \bibinfo{person}{Yanping Huang}, {and} \bibinfo{person}{Quoc~V Le}.} \bibinfo{year}{2019}\natexlab{}.
\newblock \showarticletitle{Regularized evolution for image classifier architecture search}. In \bibinfo{booktitle}{\emph{Proceedings of the aaai conference on artificial intelligence}}, Vol.~\bibinfo{volume}{33}. \bibinfo{pages}{4780--4789}.
\newblock


\bibitem[Rendle et~al\mbox{.}(2011)]%
        {rendle2011fast}
\bibfield{author}{\bibinfo{person}{Steffen Rendle}, \bibinfo{person}{Zeno Gantner}, \bibinfo{person}{Christoph Freudenthaler}, {and} \bibinfo{person}{Lars Schmidt-Thieme}.} \bibinfo{year}{2011}\natexlab{}.
\newblock \showarticletitle{Fast context-aware recommendations with factorization machines}. In \bibinfo{booktitle}{\emph{Proceedings of the 34th international ACM SIGIR conference on Research and development in Information Retrieval}}. \bibinfo{pages}{635--644}.
\newblock


\bibitem[Research(2022)]%
        {fvcore}
\bibfield{author}{\bibinfo{person}{Facebook Research}.} \bibinfo{year}{2022}\natexlab{}.
\newblock \bibinfo{title}{fvcore}.
\newblock \bibinfo{howpublished}{\url{https://github.com/facebookresearch/fvcore}}.
\newblock


\bibitem[Richardson et~al\mbox{.}(2007)]%
        {richardson2007predicting}
\bibfield{author}{\bibinfo{person}{Matthew Richardson}, \bibinfo{person}{Ewa Dominowska}, {and} \bibinfo{person}{Robert Ragno}.} \bibinfo{year}{2007}\natexlab{}.
\newblock \showarticletitle{Predicting clicks: estimating the click-through rate for new ads}. In \bibinfo{booktitle}{\emph{Proceedings of the 16th international conference on World Wide Web}}. \bibinfo{pages}{521--530}.
\newblock


\bibitem[Shan et~al\mbox{.}(2016)]%
        {shan2016deep}
\bibfield{author}{\bibinfo{person}{Ying Shan}, \bibinfo{person}{T~Ryan Hoens}, \bibinfo{person}{Jian Jiao}, \bibinfo{person}{Haijing Wang}, \bibinfo{person}{Dong Yu}, {and} \bibinfo{person}{JC Mao}.} \bibinfo{year}{2016}\natexlab{}.
\newblock \showarticletitle{Deep crossing: Web-scale modeling without manually crafted combinatorial features}. In \bibinfo{booktitle}{\emph{Proceedings of the 22nd ACM SIGKDD international conference on knowledge discovery and data mining}}. \bibinfo{pages}{255--262}.
\newblock


\bibitem[So et~al\mbox{.}(2019)]%
        {so2019evolved}
\bibfield{author}{\bibinfo{person}{David So}, \bibinfo{person}{Quoc Le}, {and} \bibinfo{person}{Chen Liang}.} \bibinfo{year}{2019}\natexlab{}.
\newblock \showarticletitle{The evolved transformer}. In \bibinfo{booktitle}{\emph{International Conference on Machine Learning}}. PMLR, \bibinfo{pages}{5877--5886}.
\newblock


\bibitem[Song et~al\mbox{.}(2020)]%
        {song2020towards}
\bibfield{author}{\bibinfo{person}{Qingquan Song}, \bibinfo{person}{Dehua Cheng}, \bibinfo{person}{Hanning Zhou}, \bibinfo{person}{Jiyan Yang}, \bibinfo{person}{Yuandong Tian}, {and} \bibinfo{person}{Xia Hu}.} \bibinfo{year}{2020}\natexlab{}.
\newblock \showarticletitle{Towards automated neural interaction discovery for click-through rate prediction}. In \bibinfo{booktitle}{\emph{Proceedings of the 26th ACM SIGKDD International Conference on Knowledge Discovery \& Data Mining}}. \bibinfo{pages}{945--955}.
\newblock


\bibitem[Song et~al\mbox{.}(2019)]%
        {song2019autoint}
\bibfield{author}{\bibinfo{person}{Weiping Song}, \bibinfo{person}{Chence Shi}, \bibinfo{person}{Zhiping Xiao}, \bibinfo{person}{Zhijian Duan}, \bibinfo{person}{Yewen Xu}, \bibinfo{person}{Ming Zhang}, {and} \bibinfo{person}{Jian Tang}.} \bibinfo{year}{2019}\natexlab{}.
\newblock \showarticletitle{Autoint: Automatic feature interaction learning via self-attentive neural networks}. In \bibinfo{booktitle}{\emph{Proceedings of the 28th ACM International Conference on Information and Knowledge Management}}. \bibinfo{pages}{1161--1170}.
\newblock


\bibitem[Sun et~al\mbox{.}(2023)]%
        {gibbon}
\bibfield{author}{\bibinfo{person}{Hanbo Sun}, \bibinfo{person}{Zhenhua Zhu}, \bibinfo{person}{Chenyu Wang}, \bibinfo{person}{Xuefei Ning}, \bibinfo{person}{Guohao Dai}, \bibinfo{person}{Huazhong Yang}, {and} \bibinfo{person}{Yu Wang}.} \bibinfo{year}{2023}\natexlab{}.
\newblock \showarticletitle{Gibbon: An Efficient Co-Exploration Framework of {NN} Model and Processing-In-Memory Architecture}.
\newblock \bibinfo{journal}{\emph{{IEEE} Trans. Comput. Aided Des. Integr. Circuits Syst.}} \bibinfo{volume}{42}, \bibinfo{number}{11} (\bibinfo{year}{2023}), \bibinfo{pages}{4075--4089}.
\newblock
\urldef\tempurl%
\url{https://doi.org/10.1109/TCAD.2023.3262201}
\showDOI{\tempurl}


\bibitem[Vaswani et~al\mbox{.}(2017)]%
        {vaswani2017attention}
\bibfield{author}{\bibinfo{person}{Ashish Vaswani}, \bibinfo{person}{Noam Shazeer}, \bibinfo{person}{Niki Parmar}, \bibinfo{person}{Jakob Uszkoreit}, \bibinfo{person}{Llion Jones}, \bibinfo{person}{Aidan~N Gomez}, \bibinfo{person}{{\L}ukasz Kaiser}, {and} \bibinfo{person}{Illia Polosukhin}.} \bibinfo{year}{2017}\natexlab{}.
\newblock \showarticletitle{Attention is all you need}.
\newblock \bibinfo{journal}{\emph{Advances in neural information processing systems}}  \bibinfo{volume}{30} (\bibinfo{year}{2017}).
\newblock


\bibitem[Wang et~al\mbox{.}(2020b)]%
        {wang2020hat}
\bibfield{author}{\bibinfo{person}{Hanrui Wang}, \bibinfo{person}{Zhanghao Wu}, \bibinfo{person}{Zhijian Liu}, \bibinfo{person}{Han Cai}, \bibinfo{person}{Ligeng Zhu}, \bibinfo{person}{Chuang Gan}, {and} \bibinfo{person}{Song Han}.} \bibinfo{year}{2020}\natexlab{b}.
\newblock \showarticletitle{Hat: Hardware-aware transformers for efficient natural language processing}.
\newblock \bibinfo{journal}{\emph{arXiv preprint arXiv:2005.14187}} (\bibinfo{year}{2020}).
\newblock


\bibitem[Wang et~al\mbox{.}(2017)]%
        {wang2017deep}
\bibfield{author}{\bibinfo{person}{Ruoxi Wang}, \bibinfo{person}{Bin Fu}, \bibinfo{person}{Gang Fu}, {and} \bibinfo{person}{Mingliang Wang}.} \bibinfo{year}{2017}\natexlab{}.
\newblock \showarticletitle{Deep \& cross network for ad click predictions}.
\newblock In \bibinfo{booktitle}{\emph{Proceedings of the ADKDD'17}}. \bibinfo{pages}{1--7}.
\newblock


\bibitem[Wang et~al\mbox{.}(2021b)]%
        {wang2021dcn}
\bibfield{author}{\bibinfo{person}{Ruoxi Wang}, \bibinfo{person}{Rakesh Shivanna}, \bibinfo{person}{Derek Cheng}, \bibinfo{person}{Sagar Jain}, \bibinfo{person}{Dong Lin}, \bibinfo{person}{Lichan Hong}, {and} \bibinfo{person}{Ed Chi}.} \bibinfo{year}{2021}\natexlab{b}.
\newblock \showarticletitle{DCN V2: Improved deep \& cross network and practical lessons for web-scale learning to rank systems}. In \bibinfo{booktitle}{\emph{Proceedings of the Web Conference 2021}}. \bibinfo{pages}{1785--1797}.
\newblock


\bibitem[Wang et~al\mbox{.}(2020a)]%
        {Wang2020APQ}
\bibfield{author}{\bibinfo{person}{Tianzhe Wang}, \bibinfo{person}{Kuan Wang}, \bibinfo{person}{Han Cai}, \bibinfo{person}{Ji Lin}, \bibinfo{person}{Zhijian Liu}, {and} \bibinfo{person}{Song Han}.} \bibinfo{year}{2020}\natexlab{a}.
\newblock \showarticletitle{APQ: Joint Search for Nerwork Architecture, Pruning and Quantization Policy}. In \bibinfo{booktitle}{\emph{Proceedings of the IEEE/CVF Conference on Computer Vision and Pattern Recognition}}.
\newblock


\bibitem[Wang et~al\mbox{.}(2021c)]%
        {wang2021rerec}
\bibfield{author}{\bibinfo{person}{Yitu Wang}, \bibinfo{person}{Zhenhua Zhu}, \bibinfo{person}{Fan Chen}, \bibinfo{person}{Mingyuan Ma}, \bibinfo{person}{Guohao Dai}, \bibinfo{person}{Yu Wang}, \bibinfo{person}{Hai Li}, {and} \bibinfo{person}{Yiran Chen}.} \bibinfo{year}{2021}\natexlab{c}.
\newblock \showarticletitle{Rerec: In-reram acceleration with access-aware mapping for personalized recommendation}. In \bibinfo{booktitle}{\emph{2021 IEEE/ACM International Conference On Computer Aided Design (ICCAD)}}. IEEE, \bibinfo{pages}{1--9}.
\newblock


\bibitem[Wang et~al\mbox{.}(2021a)]%
        {wang2021masknet}
\bibfield{author}{\bibinfo{person}{Zhiqiang Wang}, \bibinfo{person}{Qingyun She}, {and} \bibinfo{person}{Junlin Zhang}.} \bibinfo{year}{2021}\natexlab{a}.
\newblock \showarticletitle{MaskNet: introducing feature-wise multiplication to CTR ranking models by instance-guided mask}.
\newblock \bibinfo{journal}{\emph{arXiv preprint arXiv:2102.07619}} (\bibinfo{year}{2021}).
\newblock


\bibitem[Wen et~al\mbox{.}(2020)]%
        {wen2020neural}
\bibfield{author}{\bibinfo{person}{Wei Wen}, \bibinfo{person}{Hanxiao Liu}, \bibinfo{person}{Yiran Chen}, \bibinfo{person}{Hai Li}, \bibinfo{person}{Gabriel Bender}, {and} \bibinfo{person}{Pieter-Jan Kindermans}.} \bibinfo{year}{2020}\natexlab{}.
\newblock \showarticletitle{Neural predictor for neural architecture search}. In \bibinfo{booktitle}{\emph{European Conference on Computer Vision}}. Springer, \bibinfo{pages}{660--676}.
\newblock


\bibitem[Yan and Li(2023)]%
        {yan2023xdeepint}
\bibfield{author}{\bibinfo{person}{Yachen Yan} {and} \bibinfo{person}{Liubo Li}.} \bibinfo{year}{2023}\natexlab{}.
\newblock \showarticletitle{xDeepInt: a hybrid architecture for modeling the vector-wise and bit-wise feature interactions}.
\newblock \bibinfo{journal}{\emph{arXiv preprint arXiv:2301.01089}} (\bibinfo{year}{2023}).
\newblock


\bibitem[Yang et~al\mbox{.}(2022)]%
        {yang2022research}
\bibfield{author}{\bibinfo{person}{Xiaoxuan Yang} {et~al\mbox{.}}} \bibinfo{year}{2022}\natexlab{}.
\newblock \showarticletitle{Research progress on memristor: From synapses to computing systems}.
\newblock \bibinfo{journal}{\emph{IEEE Transactions on Circuits and Systems I: Regular Papers}} \bibinfo{volume}{69}, \bibinfo{number}{5} (\bibinfo{year}{2022}), \bibinfo{pages}{1845--1857}.
\newblock


\bibitem[Yu et~al\mbox{.}(2020)]%
        {yu2020bignas}
\bibfield{author}{\bibinfo{person}{Jiahui Yu}, \bibinfo{person}{Pengchong Jin}, \bibinfo{person}{Hanxiao Liu}, \bibinfo{person}{Gabriel Bender}, \bibinfo{person}{Pieter-Jan Kindermans}, \bibinfo{person}{Mingxing Tan}, \bibinfo{person}{Thomas Huang}, \bibinfo{person}{Xiaodan Song}, \bibinfo{person}{Ruoming Pang}, {and} \bibinfo{person}{Quoc Le}.} \bibinfo{year}{2020}\natexlab{}.
\newblock \showarticletitle{Bignas: Scaling up neural architecture search with big single-stage models}. In \bibinfo{booktitle}{\emph{European Conference on Computer Vision}}. Springer, \bibinfo{pages}{702--717}.
\newblock


\bibitem[Zhang et~al\mbox{.}(2023)]%
        {zhang2023nasrec}
\bibfield{author}{\bibinfo{person}{Tunhou Zhang}, \bibinfo{person}{Dehua Cheng}, \bibinfo{person}{Yuchen He}, \bibinfo{person}{Zhengxing Chen}, \bibinfo{person}{Xiaoliang Dai}, \bibinfo{person}{Liang Xiong}, \bibinfo{person}{Feng Yan}, \bibinfo{person}{Hai Li}, \bibinfo{person}{Yiran Chen}, {and} \bibinfo{person}{Wei Wen}.} \bibinfo{year}{2023}\natexlab{}.
\newblock \showarticletitle{NASRec: weight sharing neural architecture search for recommender systems}. In \bibinfo{booktitle}{\emph{Proceedings of the ACM Web Conference 2023}}. \bibinfo{pages}{1199--1207}.
\newblock


\bibitem[Zhou et~al\mbox{.}(2018)]%
        {zhou2018deep}
\bibfield{author}{\bibinfo{person}{Guorui Zhou}, \bibinfo{person}{Xiaoqiang Zhu}, \bibinfo{person}{Chenru Song}, \bibinfo{person}{Ying Fan}, \bibinfo{person}{Han Zhu}, \bibinfo{person}{Xiao Ma}, \bibinfo{person}{Yanghui Yan}, \bibinfo{person}{Junqi Jin}, \bibinfo{person}{Han Li}, {and} \bibinfo{person}{Kun Gai}.} \bibinfo{year}{2018}\natexlab{}.
\newblock \showarticletitle{Deep interest network for click-through rate prediction}. In \bibinfo{booktitle}{\emph{Proceedings of the 24th ACM SIGKDD international conference on knowledge discovery \& data mining}}. \bibinfo{pages}{1059--1068}.
\newblock


\bibitem[Zhu et~al\mbox{.}(2020)]%
        {mnsim}
\bibfield{author}{\bibinfo{person}{Zhenhua Zhu}, \bibinfo{person}{Hanbo Sun}, \bibinfo{person}{Kaizhong Qiu}, \bibinfo{person}{Lixue Xia}, \bibinfo{person}{Gokul Krishnan}, \bibinfo{person}{Guohao Dai}, \bibinfo{person}{Dimin Niu}, \bibinfo{person}{Xiaoming Chen}, \bibinfo{person}{Xiaobo~Sharon Hu}, \bibinfo{person}{Yu Cao}, \bibinfo{person}{Yuan Xie}, \bibinfo{person}{Yu Wang}, {and} \bibinfo{person}{Huazhong Yang}.} \bibinfo{year}{2020}\natexlab{}.
\newblock \showarticletitle{{MNSIM} 2.0: {A} Behavior-Level Modeling Tool for Memristor-based Neuromorphic Computing Systems}. In \bibinfo{booktitle}{\emph{{GLSVLSI} '20: Great Lakes Symposium on {VLSI} 2020, Virtual Event, China, September 7-9, 2020}}, \bibfield{editor}{\bibinfo{person}{Tinoosh Mohsenin}, \bibinfo{person}{Weisheng Zhao}, \bibinfo{person}{Yiran Chen}, {and} \bibinfo{person}{Onur Mutlu}} (Eds.). \bibinfo{publisher}{{ACM}}, \bibinfo{pages}{83--88}.
\newblock


\bibitem[Zoph et~al\mbox{.}(2018)]%
        {zoph2018learning}
\bibfield{author}{\bibinfo{person}{Barret Zoph}, \bibinfo{person}{Vijay Vasudevan}, \bibinfo{person}{Jonathon Shlens}, {and} \bibinfo{person}{Quoc~V Le}.} \bibinfo{year}{2018}\natexlab{}.
\newblock \showarticletitle{Learning transferable architectures for scalable image recognition}. In \bibinfo{booktitle}{\emph{Proceedings of the IEEE conference on computer vision and pattern recognition}}. \bibinfo{pages}{8697--8710}.
\newblock


\end{thebibliography}
